\documentclass[11pt]{article} 

\usepackage{amssymb}
\usepackage{amsmath}
\usepackage{url}
\usepackage{lineno}
\usepackage{soul}
\usepackage{lscape}
\usepackage{rotating}
\usepackage{makecell}
\usepackage[utf8]{inputenc}
\usepackage[T1]{fontenc}
\usepackage{lmodern}
\usepackage{graphicx}
\usepackage{float}
\usepackage[figurename=Fig.,labelfont=bf,labelsep=period]{caption}
\usepackage{subcaption}
\usepackage{xspace}
\usepackage{microtype}
\usepackage{xcolor}
\usepackage{colortbl}
\usepackage{multicol}
\usepackage[left=1in,right=1in,top=1in,bottom=1in]{geometry}
\usepackage[noblocks]{authblk}
\usepackage{xr}
\begin{document}

\title{Finite Element and Machine Learning Modeling of Autogenous Self-Healing Concrete}

\author{William Liu$^{1}$\\
$^1$Pennsylvania State University, University Park, Pennsylvania, USA 16802.\\
\texttt{liuwilliam12@gmail.com}\\
}

\date{}

\maketitle

\begin{abstract}
A time-dependent modeling framework for autogenous self-healing concrete that couples moisture diffusion with damage evolution was developed. Water transport follows Fick's second law with a damage-dependent diffusivity obtained by power-law interpolation between intact concrete and crack space. Healing reduces damage in proportion to the product of local moisture and a smoothed cement availability field computed via a novel Helmholtz filtering approach that models the spatial extent over which cement clinker can travel and form crystals. Two finite element variants were implemented in FEniCSx over time horizons up to $5\times10^6$ seconds: a Crack Diffusion Model (CDM) with standard diffusion and a Crack Membrane Model (CMM) that introduces a novel threshold-based gating mechanism to control cross-crack water transport until a critical moisture threshold is reached. Key control parameters are the initial crack orientation and size, the diffusion coefficients of intact and cracked concrete, the healing rate constant, and the cement availability smoothing parameter. Simulations show that healing time varies non-monotonically with crack orientation, peaking near $45^\circ$ and $135^\circ$ and minimizing near $90^\circ$, consistent with diffusion distance to crack endpoints dominating the process. The dependence on crack width reverses with material parameters: healing time increases when $D_{\text{cracked}}<D_{\text{intact}}$ and decreases when $D_{\text{cracked}}>D_{\text{intact}}$. The CMM reproduces staged moisture penetration with delayed gate activation but lengthens total healing time, whereas the CDM is efficient for parametric sweeps. Machine learning regression models were trained on finite element simulation data to predict healing times $H(\sigma,\gamma,\beta)$ with high accuracy ($R^2 > 0.999$), dramatically reducing computational time. SHAP analysis demonstrated that crack orientation influenced healing time the most, followed by crack width and cement availability smoothing. Although experimental calibration is still required, the framework provides a versatile tool for guiding future laboratory studies and implementations of self-healing concrete.
\end{abstract}


\section{Introduction}\label{sec:introduction}
\subsection{Background}\label{sec:background}
Concrete is the second most widely used material globally, surpassed only by water, with over 30 billion metric tons utilized worldwide annually \cite{nguyen2023review}. This critical construction material has been employed for over 8,000 years, dating back to concrete-like structures built by Nabataean traders in modern-day Syria \cite{nachi2025history}, and continues to be essential for housing, transportation, and water infrastructure today. The global concrete market is projected to reach \$821.6 billion by 2026 \cite{gitnux2025concrete}, driven by growing investment in construction projects worldwide.

However, concrete production and usage come with significant environmental and economic costs. The concrete industry contributes approximately 8\% of global CO$_2$ emissions, exceeding the combined emissions from the aviation (3.5\%) and shipping (3\%) industries \cite{andrewGlobalCO2Emissions2024}. Under current trajectories, carbon emissions from concrete production are expected to increase from 1.6 billion tonnes today to 3.8 billion tonnes annually by 2050 \cite{weforum2024cement}. To meet the Paris Climate Agreement's 1.5$^\circ$C warming limit, the global concrete industry must reduce emissions by 16\% by 2030 and achieve net-zero emissions by 2050, while simultaneously addressing the projected 48\% increase in cement demand from 4.2 billion to 6.2 billion tons by 2050 \cite{rmi2023concrete}.

The infrastructure crisis compounds these environmental challenges. Most countries completed their major infrastructure projects during the late 1900s, including the US Interstate Highway System, European motorways, Japan's Shinkansen, and China's expressway networks. Over half a century later, these aging structures face increasing failure risks, with concrete failure probability approaching 70\% after 50 years and becoming nearly inevitable by the century's end \cite{structuremag2016durability}. Currently, over 20,000 concrete bridges in the United States are classified as structurally deficient \cite{asce2025infrastructure}, requiring frequent maintenance and posing safety risks.

Crack formation represents an inherent characteristic of concrete structures, arising from multiple mechanisms: structural loading that exceeds concrete's limited tensile strength, volumetric changes associated with shrinkage and expansion, and environmental degradation processes including corrosion and freeze-thaw cycles \cite{han2017smart,zhang2020selfhealing}. The presence of cracks significantly compromises concrete durability by providing pathways for liquids and gases to penetrate to the steel reinforcement, thereby accelerating corrosion processes. The global cost of corrosion in reinforced concrete infrastructure is estimated at \$2.5 trillion annually, representing approximately 3.4\% of global GDP \cite{nace2025corrosion}.

Self-healing concrete emerges as a promising solution to address these challenges. This innovative material repairs its cracks automatically without human intervention, typically through chemical or biological processes \cite{hearnSelfsealingAutogenousHealing1998}. Self-healing concrete offers reduced maintenance requirements over the infrastructure lifetime, diminished social costs of reconstruction, increased infrastructure efficiency, decreased economic costs of frequent repairs, and reduced environmental impacts of regular restorations \cite{edvardsenWaterPermeabilityAutogenous1999a}.

Self-healing concrete approaches are broadly categorized into two types: autogenous healing, where water penetration triggers intrinsic chemical healing reactions, and autonomous healing, which relies on discrete self-healing mechanisms added during production \cite{hearnSelfsealingAutogenousHealing1998}. While autonomous healing can address larger cracks (0.1-1.8 mm), it faces challenges including low survivability during mixing, harsh biological environments, and high cost \cite{amran2022selfhealing,usc2022bacteria}.

For these reasons, this study focuses on autogenous self-healing concrete, which more closely resembles traditional concrete and offers greater potential for real-world implementation. Autogenous self-healing concrete exhibits superior computational tractability compared to bacteria-based systems due to its uniformity and predictability. All concrete types possess inherent autogenous self-healing capabilities through continued hydration, carbonation, and calcium silicate hydrate (C-S-H) swelling, although the healed material generally exhibits inferior properties compared to the original material \cite{han2017smart,zhang2020selfhealing}. For most multi-mineral materials, optimal self-healing performance is achieved through a combination of silica-based swelling and crystalline components, enabling crack repair up to 0.6 millimeters in width \cite{zhang2020selfhealing,lahmann2022autogenous}. In this study, autogenous self-healing concrete is defined as a cementitious mixture specifically modified to enhance self-healing capacity.

Despite comprehensive laboratory investigations, limited real-world applications of autogenous self-healing concrete exist. The Cardiff-led Materials For Life (M4L) project conducted field trials testing six different concrete walls in Wales \cite{bbc2015selfhealing}. However, to the best of our knowledge, no dedicated implementation of autogenous self-healing concrete in civilian infrastructure has been documented.

Traditional laboratory studies typically focus on isolated environmental stressors, neglecting the compounded effects of multiple simultaneous stressors. Computer simulations offer a solution by providing cheaper, faster, and more consistent results than laboratory studies. Simulation enables modeling of multiple factors simultaneously, while laboratory equipment constraints often limit studies to only one or two stressors. Furthermore, large-scale infrastructure trials are rare and expensive, making simulation models valuable precursors to further investment.

\subsection{Framework and Novelty of the Study}\label{sec:framework_and_novelty}
We developed two distinct model versions using the FEniCSx finite element software \cite{fenics}: the Crack Diffusion Model (CDM) implementing standard diffusion across crack interfaces, and the Crack Membrane Model (CMM) implementing a binary gate that selectively allows diffusion only after reaching critical moisture levels. The models incorporate time-dependent feedback mechanisms, realistic diffusion coefficients, and damage evolution that couples cement and water availability.

Previous research has established a foundation for finite element analysis and physics-based modeling of autogenous self-healing concrete \cite{hilloulin2014autogenous,vedrtnam2025bacterial,shahsavari2016damage,chen2021hydration}. Building upon this foundation, the present work addresses several areas where enhanced modeling capabilities could provide more accurate predictions.

First, while some studies have employed thermodynamic approaches where healing is governed by strain and energy equivalence \cite{shahsavari2016damage,pan2018damage}, autogenous self-healing concrete is driven by the hydration of cement clinker and the diffusion of water into the crack space. A diffusion-based approach captures the chemical processes that form healing crystals and accounts for the limiting factors of water and cement clinker availability. The models implement damage-dependent anisotropic diffusion, where the diffusion coefficient varies spatially based on the local damage field. Additionally, while some previous studies have assumed continuous water transport across crack interfaces \cite{hilloulin2014autogenous}, the CMM incorporates a gating mechanism that requires sufficient moisture buildup before allowing cross-crack transport, representing the physical process of pressure-based water penetration.

The healing reactions depend on both the availability of water and cement clinker. While some studies have assumed uniform healing rates or simplified kinetics \cite{alkhuzai2023numerical,freeman2019simulation,mauludin2021computational}, the CDM and CMM incorporate a kinetics relationship that captures both water and cement clinker availability. The cement availability smoothing parameter accounts for both the spatial distance over which cement can be transported and its non-uniform, damage-dependent availability.

The models fully couple diffusion and reaction processes within a single time-dependent framework, whereas previous studies have typically implemented these processes independently \cite{kanellopoulos2022selfhealing}. By updating each equation at each time step, the models capture feedback mechanisms that combine various physical processes into a unified simulation. This fully coupled approach represents the interdependent nature of moisture transport, cement availability, and healing kinetics. To address the computational cost of the coupled models, machine learning is used to predict healing times efficiently \cite{badarinath2021machine}. Explainable AI (XAI) is used to further interpret the models' predictions.

The remainder of this paper is organized as follows: Section \ref{sec:modeling_approach} details the development of the Crack Diffusion Model, including governing equations, variables, and implementation methodology. Section \ref{sec:crack_membrane_model} introduces the Crack Membrane Model as an enhanced approach. Section \ref{sec:machine_learning_model} demonstrates machine learning applications for predicting healing behavior and reducing computational costs. Section \ref{sec:results_and_analysis} presents results and analysis of healing time relationships with crack angle and width parameters, including performance comparisons between CDM and CMM models and machine learning model evaluations. Section \ref{sec:limitations_and_future_extensions} discusses limitations and future extensions, and Section \ref{sec:conclusions} concludes with key findings and implications for infrastructure sustainability.

\section{Modeling Approach}\label{sec:modeling_approach}

\subsection{Crack Diffusion Model}\label{sec:crack_diffusion_model}
CDM simulates autogenous healing through moisture transport and cement hydration processes. The model couples moisture diffusion with damage evolution, accounting for the spatial heterogeneity of concrete microstructure and the availability of unhydrated cement clinker.

Water diffusion follows Fick's second law with a damage-dependent diffusivity:
\begin{equation}
\frac{\partial U}{\partial t} = \nabla \cdot (D(d)\nabla U)
\label{eq:ficks_second_law}
\end{equation}
where $U$ is the water concentration variable (dimensionless), $D(d)$ is the diffusion coefficient ($\frac{\text{cm}^2}{\text{s}}$) as a function of $d$, and $d$ is the damage variable (dimensionless). The damage variable represents the fraction of damaged area, where $d = 0$ corresponds to intact concrete and $d = 1$ to complete damage.

The effective diffusion coefficient interpolates between intact concrete ($D_{\text{intact}}$) and crack space ($D_{\text{cracked}}$) using a power-law relationship:
\begin{equation}
D(d) = D_{\text{intact}}^{(1-d)^p} \cdot D_{\text{cracked}}^{(1-(1-d)^p)}
\label{eq:diffusion_coefficient}
\end{equation}
where $p$ controls the sharpness of the transition. The crack-space diffusivity accounts for microcrack bridging, capillary effects, and tortuous percolation pathways.

Healing reduces damage through cement hydration, which is modeled as:
\begin{equation}
\frac{\partial d}{\partial t} = -\alpha U \cdot \chi_{\text{eff}}(d, q)
\label{eq:damage_update}
\end{equation}
where $\alpha$ is the healing rate constant and $\chi_{\text{eff}}$ represents effective cement availability. The local cement availability follows:
\begin{equation}
\chi(d, q) = (1 - d)^q
\label{eq:chi_function}
\end{equation}
where $q$ controls the nonlinear decrease in cement availability with damage.

A complete crack introduces a discontinuity in the cement availability field. To enable healing modeling in completely damaged regions ($d = 1$), this discontinuity must be smoothed using Helmholtz filtering:
\begin{align}
\chi_{\text{eff}}(x) - \gamma \Delta \chi_{\text{eff}}(x) &= \chi(x), \quad \forall x \in \Omega \notag \\
\nabla \chi_{\text{eff}} \cdot n &= 0, \quad \forall x \in \partial \Omega \label{eq:helmholtz_pde_neumann}
\end{align}
where the spatial domain and its boundary are denoted as $\Omega$ and $\partial \Omega$, respectively, and $\gamma$ is the characteristic smoothing length scale. The smoothing parameter describes the distance over which cement clinker can be transported by water to form healing crystals.

\begin{figure}[htp]
    \centering
    \includegraphics[width=1\linewidth]{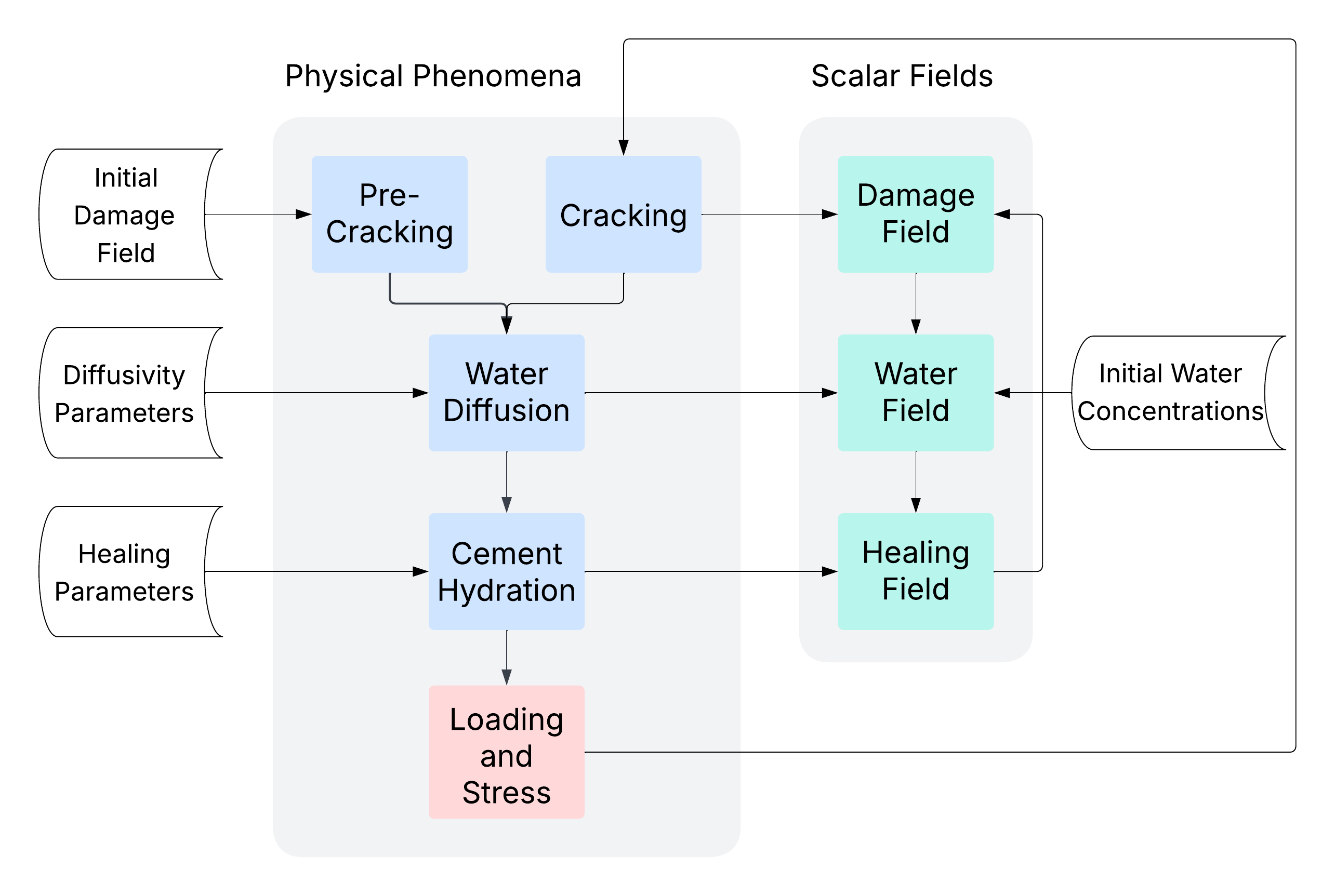}
    \caption{Autogenous Self-Healing Concrete Simulation Workflow using CDM with Parameters, Phenomena, and Scalar Fields}
    \label{fig:model_workflow}
\end{figure}

The CDM solves the coupled system of equations introduced above using the latest version of FEniCSx in Python, with the workflow shown in Figure~\ref{fig:model_workflow}. The time-dependent solution accounts for feedback between moisture transport, damage evolution, and cement availability. 

\subsection{Crack Membrane Model}\label{sec:crack_membrane_model}

CMM addresses a key limitation of CDM: the instantaneous water transport across cracks. In reality, water encounters resistance at crack boundaries and requires sufficient pressure to penetrate crack spaces. The CMM implements a threshold-based gating mechanism that controls water flow across damaged regions.

Specifically, CMM introduces a damage threshold $d_{\text{threshold}}$ that distinguishes intact regions ($d \leq d_{\text{threshold}}$) from cracked regions ($d > d_{\text{threshold}}$). Water transport in cracked regions is governed by a gating function $G(U)$ that depends on local moisture concentration:

\begin{equation}
G(U) = \epsilon_{\text{gate}} + \frac{1 - \epsilon_{\text{gate}}}{1 + \exp(-(U - U_{\text{critical}})/\Delta U_{\text{gate}})}
\label{eq:gate_value}
\end{equation}
where $U_{\text{critical}}$ is the critical moisture threshold, $\Delta U_{\text{gate}}$ controls the transition sharpness, and $\epsilon_{\text{gate}}$ is a small positive constant that prevents numerical instabilities. When $U>>U_{\text{critical}}$, $G(U) = 1$, allowing for normal diffusion. When $U<<U_{\text{critical}}$, $G(U) = 0$, preventing water transport.

In CMM, the effective diffusion coefficient $D_{\text{eff}}$ becomes a function of both damage and moisture concentration:
\begin{equation}
D_{\text{eff}}(d, U) = \begin{cases}
D(d) & \text{if } d \leq d_{\text{threshold}} \\
G(U) \cdot D(d) & \text{if } d > d_{\text{threshold}}
\end{cases}
\end{equation}

\begin{figure}[htp]
    \centering
    \includegraphics[width=0.8\linewidth]{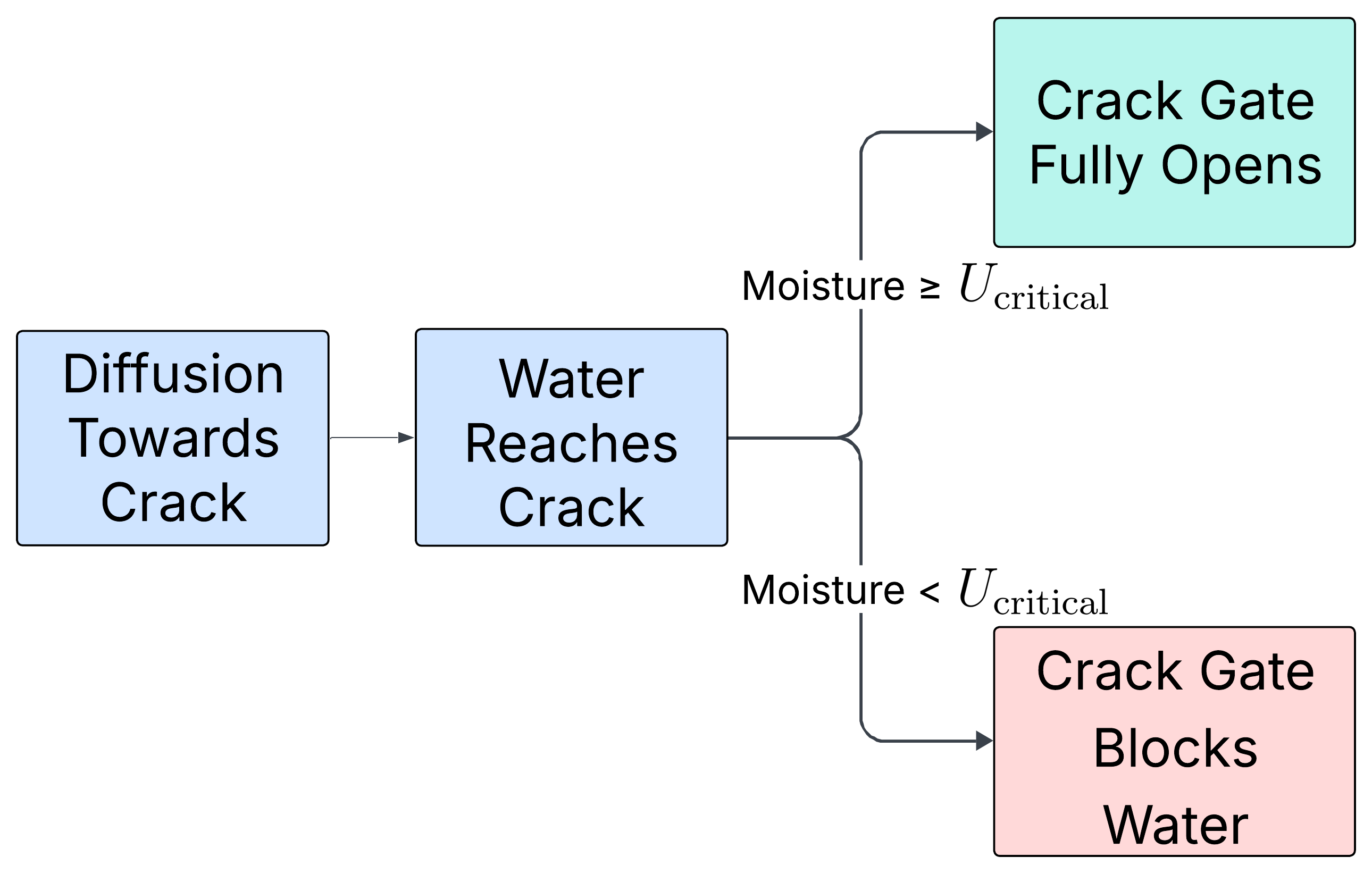}
    \caption{Crack Membrane Model Logic Workflow for Water Transport Across Cracks}
    \label{fig:crack_membrane_workflow}
\end{figure}

This formulation suppresses water flow in cracked regions until the critical moisture threshold and water pressure are reached, after which normal diffusion across the crack resumes. The CMM also solves the coupled system of equations using the latest version of FEniCSx in Python, with the workflow shown in Figure~\ref{fig:crack_membrane_workflow}. 

\subsection{Machine Learning Model}\label{sec:machine_learning_model}

Simulations in this study require solving millions of partial differential equations on fine meshes, demanding substantial computational resources. Machine learning models trained on physics-based simulation data can learn underlying physical relationships and provide predictions 330 to 2,960,000 times faster than standalone finite element analysis \cite{nath2020mlfea}.

\begin{figure}[htp]
    \centering
    \includegraphics[width=1\linewidth]{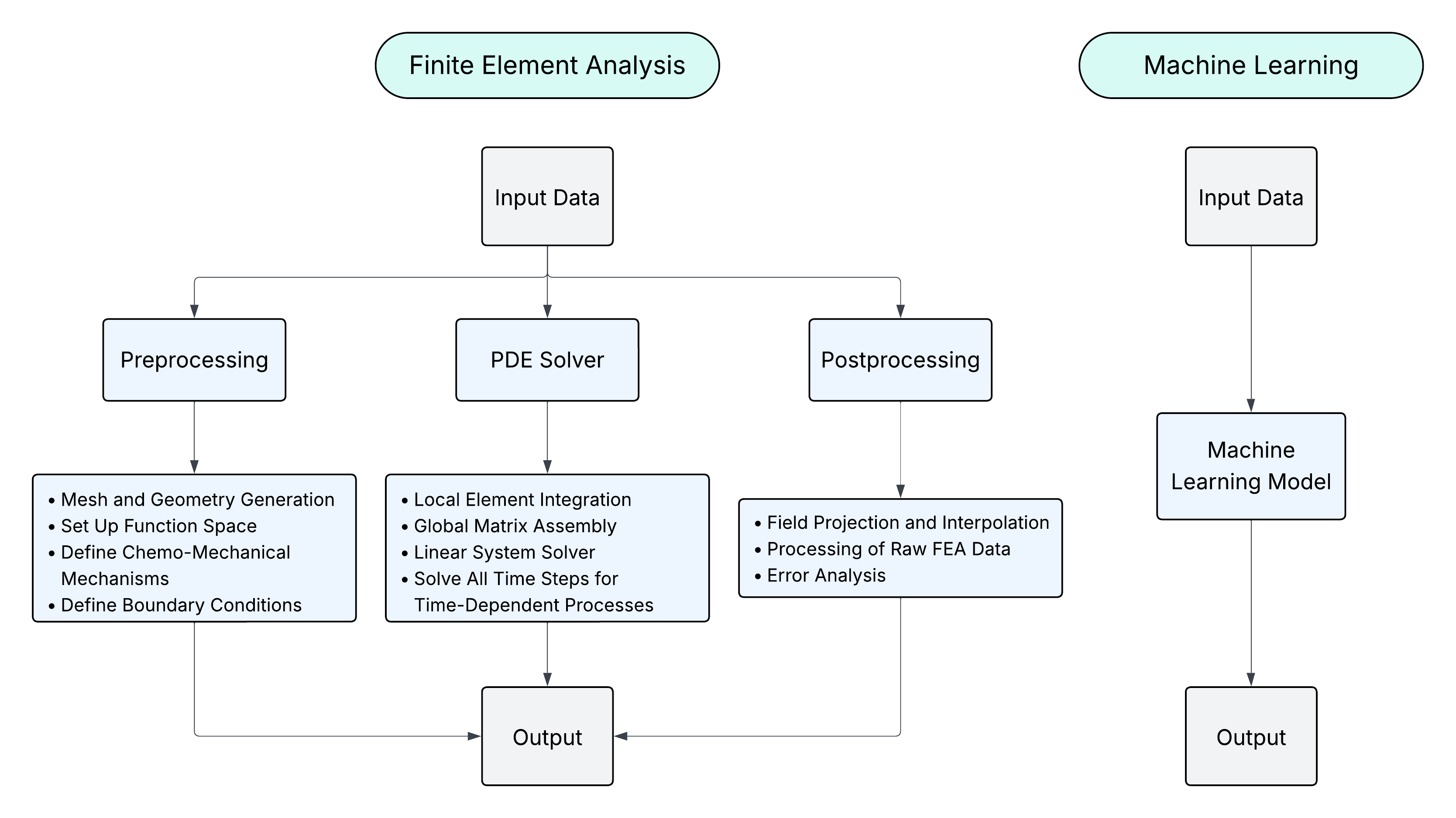}
    \caption{Comparison of the Computing Complexities between Finite Element Analysis and Machine Learning}
    \label{fig:fea_vs_ml}
\end{figure}

Figure~\ref{fig:fea_vs_ml} shows a comparison of computational requirements between physics-based FEA and trained ML models. FEA requires multiple computational stages including mesh generation, PDE discretization, linear system assembly, iterative solution, and postprocessing, with the sparse linear system solver representing the most computationally intensive component. Conversely, ML inference requires only input data preprocessing and model evaluation, generating output predictions within milliseconds. However, it should be noted that ML model training requires substantial computational resources for training data generation through systematic FEA simulations and iterative parameter optimization via backpropagation.

The training of the ML models proceeded as follows. A dataset of 1100 simulation samples was generated from the CDM by systematically varying the crack width indicator $\sigma$, cement availability smoothing length $\gamma$, and crack orientation $\beta$ over the parameter ranges detailed in Table~\ref{tab:ml_constants}. All parameters were normalized to zero mean and unit variance using z-score normalization. For each parameter combination, the healing time $H$ was computed as the time required to heal 95\% of the damage. Other parameters were kept constant. Specifically, their values were $\alpha$ = 0.01, $D_{\text{intact}}$ = $10^{-8}\frac{\text{cm}^2}{\text{s}}$, and $D_{\text{cracked}}$ = $10^{-7}\frac{\text{cm}^2}{\text{s}}$. The dataset was partitioned into training (60\%), validation (20\%), and test (20\%) subsets for model development and evaluation. Five standard regression models were trained to learn the regression function $H(\sigma, \gamma, \beta)$. Namely, they were the Gaussian Process, Ridge Regression, Random Forest, eXtreme Gradient Boosting (XGBoost), and Neural Network Models. All of the models were trained out-of-the-box with the default parameters except for the neural network. For the neural network, a 3-32-32-1 architecture with ReLU activation, dropout=0.2, learning rate=0.001, batch size=32, and 100 epochs was used. To interpret model predictions and quantify feature importance, SHapley Additive exPlanations (SHAP) values were computed for each input feature. The feature contributions are visualized using both bar charts and summary plots. Implementation utilized the latest releases of the TensorFlow, scikit-learn, XGBoost, and SHAP libraries in Python.

\begin{table}[htp]
    \centering
    \caption{Parameter Value Ranges and Sampling Units for ML Training Data Generation}
    \label{tab:ml_constants}    
    \begin{tabular}{l|l|l|l|l}
        \hline
        \textbf{Parameter} & \textbf{Lower Bound} & \textbf{Upper Bound} & \textbf{Step Size} & \textbf{Number of Steps} \\
        \hline
        $\sigma$ & $0.005 \text{ cm}$ & $0.05 \text{ cm}$ & $0.005 \text{ cm}$ & $10$ \\
        \hline
        $\gamma$ & $0.005 \text{ cm}$ & $0.05 \text{ cm}$ & $0.005 \text{ cm}$ & $10$ \\
        \hline
        $\beta$ & $0$ & $\frac{\pi}{2}$ & $\frac{\pi}{20}$ & $11$ \\
        \hline
    \end{tabular}
\end{table} 

\subsection{Simulation Cases and Preparation}
\begin{figure}[htp]
    \centering    
    \includegraphics[width=\linewidth]{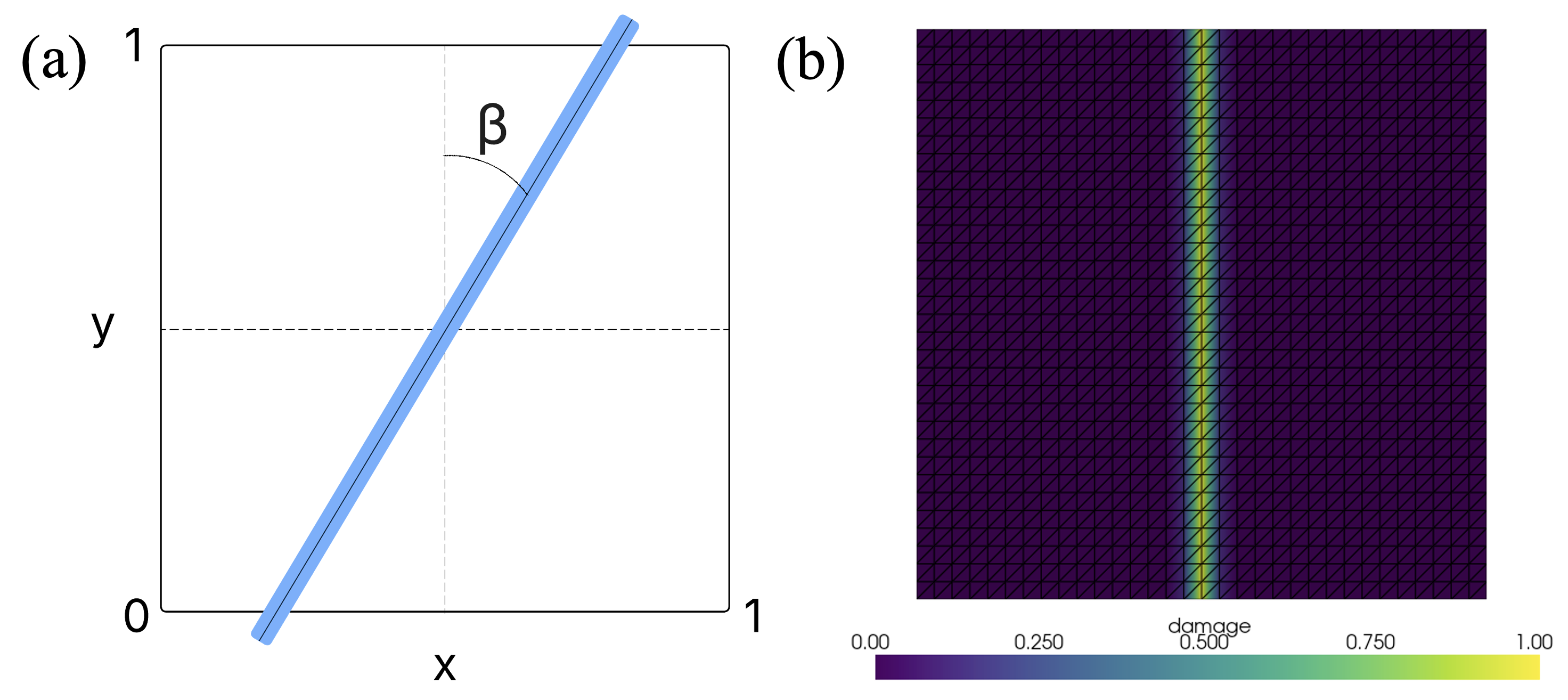}
    \caption{Initial Condition for the Damage Field: (a) Definition of Inclination Angle $\beta$, (b) Example Scalar Damage Field During Pre-Cracking ($\beta=0$, $\sigma^2 = 0.0005$)}
    \label{fig:example_vertical_crack_field}
\end{figure}

This study used a 2D, $1\times1$ cm unit square mesh discretized into 2,048 triangles of equal size. The initial condition was a pre-defined line crack passing through the center of the domain with inclination angle $\beta$ (Figure~\ref{fig:example_vertical_crack_field}(a)). The impact of the crack angle on various properties and healing results was investigated. To initialize a continuous damage field, the baseline damage follows a Gaussian function as
\begin{equation}
    d(x, y) = e^{-\frac{((x - 0.5)\cos\beta - (y - 0.5)\sin\beta)^2}{\sigma^2}}
    \label{eq:angled_damage_field}
\end{equation}
where $\sigma$ controls the width and smoothness of the Gaussian function. As an example, Figure~\ref{fig:example_vertical_crack_field}(b) shows the initial damage field for the crack with parameter values $\beta$ = 0 and $\sigma = \sqrt{0.0005} \approx 0.02236$ (99.7\% of the crack spans about 0.0671 units, or $3\sigma$).

To solve the diffusion equation, proper boundary conditions were prescribed. On the left side boundary, water supply was always available, i.e., the Dirichlet boundary condition of $U$ = 1 was established. All other boundaries acted as completely sealed barriers with homogeneous Neumann conditions, where the normal gradient of $U$ was set to zero. 

In this work, the default numerical schemes in FEniCSx were used for the solution of the nonlinear diffusion equation. The Galerkin method with continuous Lagrange basis functions was used for the trial and test spaces, leading to a variational formulation of the diffusion operator that was assembled into sparse matrices. The temporal discretization was handled by the backward Euler (implicit Euler) scheme, which is first-order accurate, unconditionally stable, and widely used as the default choice for nonlinear parabolic PDEs.

One of the main goals of this study was to uncover the influence of physical parameters on relevant concrete recovery processes. Specifically, this study addressed the following three questions:
\begin{enumerate}    
        \item What is the influence of crack angle, crack width, and cement availability on healing time?
        \item How do the healing progressions of CDM and CMM compare?
        \item How accurately can machine learning be used to predict healing, while reducing computation time?
\end{enumerate}

The models incorporate multiple interdependent parameters. Table~\ref{tab:parameters} lists the primary parameters and material constants.

\begin{table}[htp]
    \centering
    \caption{Model Parameters and Constants}
    \label{tab:parameters}
    \begin{tabular}{p{2cm}|p{7cm}|p{5cm}}
    \hline
    \textbf{Variable} & \textbf{Description} & \textbf{Constant or Parameter}\\
    \hline
    $D_{\text{intact}}$ & Diffusivity of water in pure, healed concrete & Constant: $10^{-8}\frac{\text{cm}^2}{\text{s}}$\\
    \hline
    $D_{\text{cracked}}$ & Diffusivity of water in crack space, representing microcrack bridging, capillary bridges, partially healed C-S-H gel phases, and tortuous percolation & Constant: $10^{-7}\frac{\text{cm}^2}{\text{s}}$ (except for the analysis of crack width on healing time)\\
    \hline
    $\alpha$ & Healing rate constant of proportionality & Parameter \\
    \hline
    $\beta$ & Angle of crack & Parameter \\
    \hline
    $\sigma$ & Crack width indicator & Parameter \\
    \hline
    $\gamma$ & Cement availability smoothing parameter & Parameter \\
    \hline
    $p,q$ & Sharpness variables for diffusivity and cement availability, respectively & Constant; both set to 1 \\
    \hline
    \end{tabular}
    \end{table}

\begin{figure}[htp]
    \centering
    \includegraphics[width=0.5\linewidth]{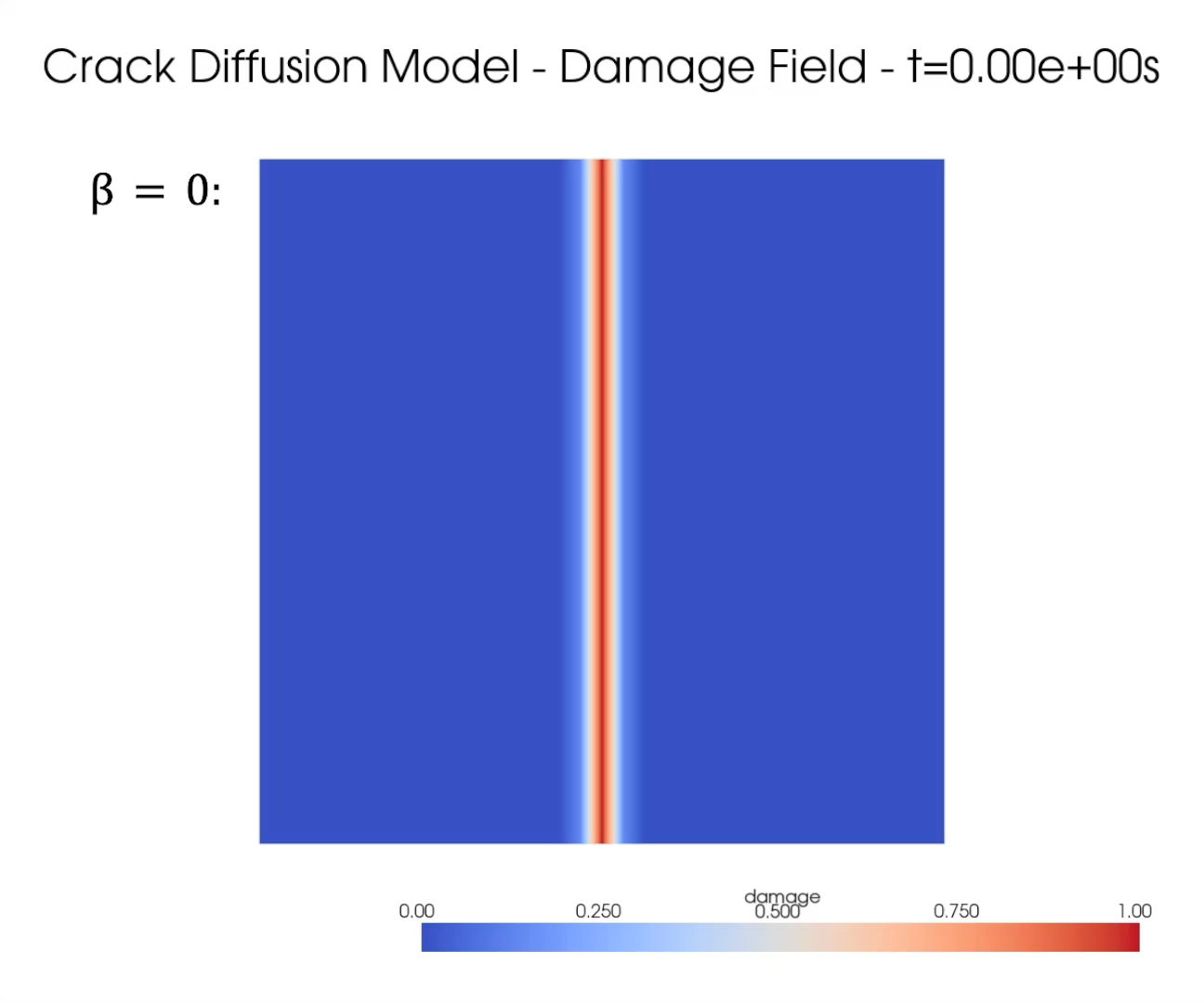}
    \caption{Video 1. CDM Damage Evolution Animations for $\beta=0, \frac{\pi}{8}, \frac{\pi}{2}$}
    \label{fig:animations1}
\end{figure}

\begin{figure}[htp]
    \centering
    \includegraphics[width=0.5\linewidth]{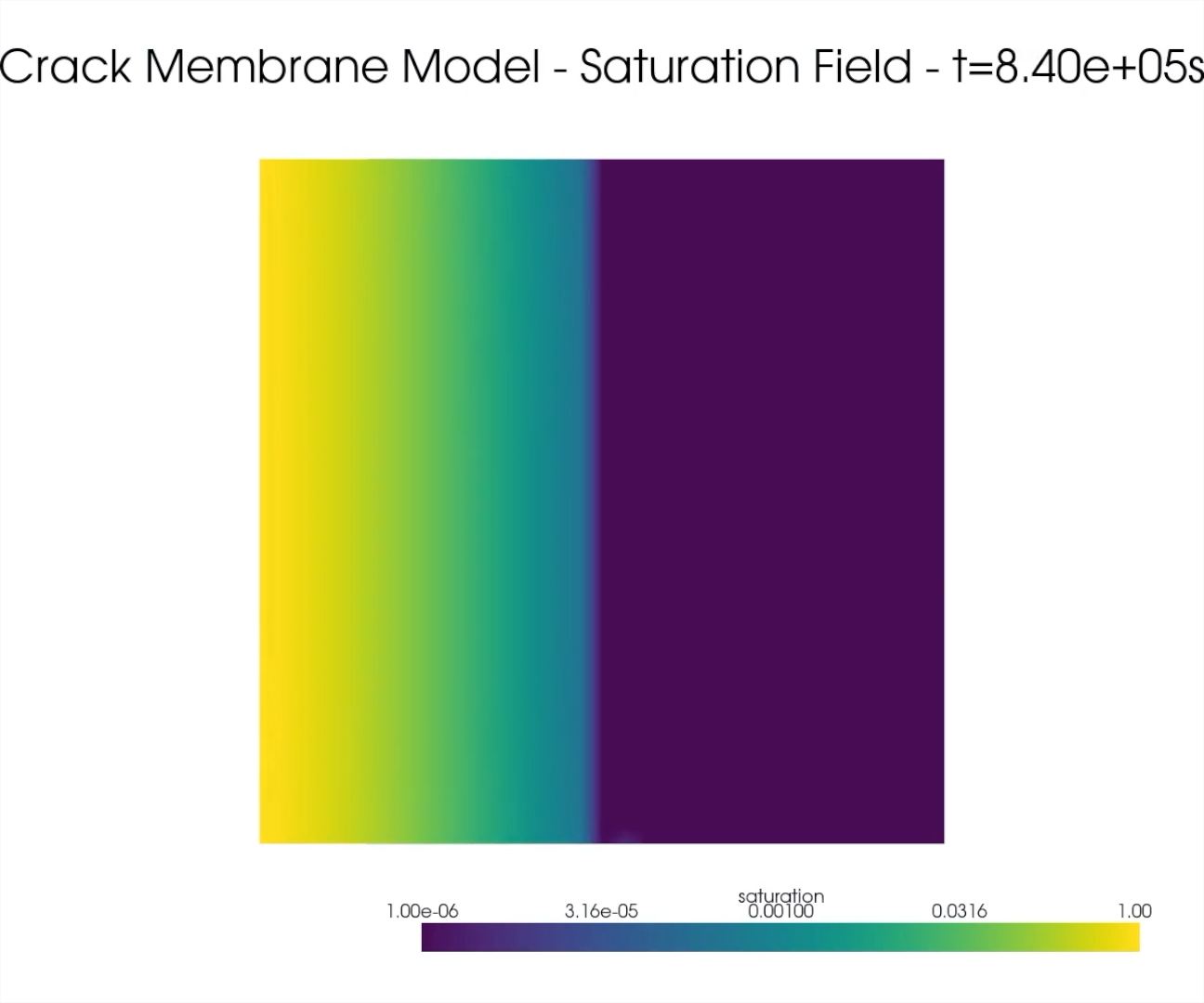}
    \caption{Video 2. CMM Damage and Saturation Evolution Animations}
    \label{fig:animations2}
\end{figure}

Animation videos were generated in Pyvista, with CDM damage evolution for various crack angles shown in Figure~\ref{fig:animations1} and CMM damage and saturation evolutions shown in Figure~\ref{fig:animations2}.
    
\section{Results and Analysis}\label{sec:results_and_analysis}

\subsection{Healing Time vs. Crack Angle}
\begin{figure}[htp]
    \centering
    \includegraphics[width=1\linewidth]{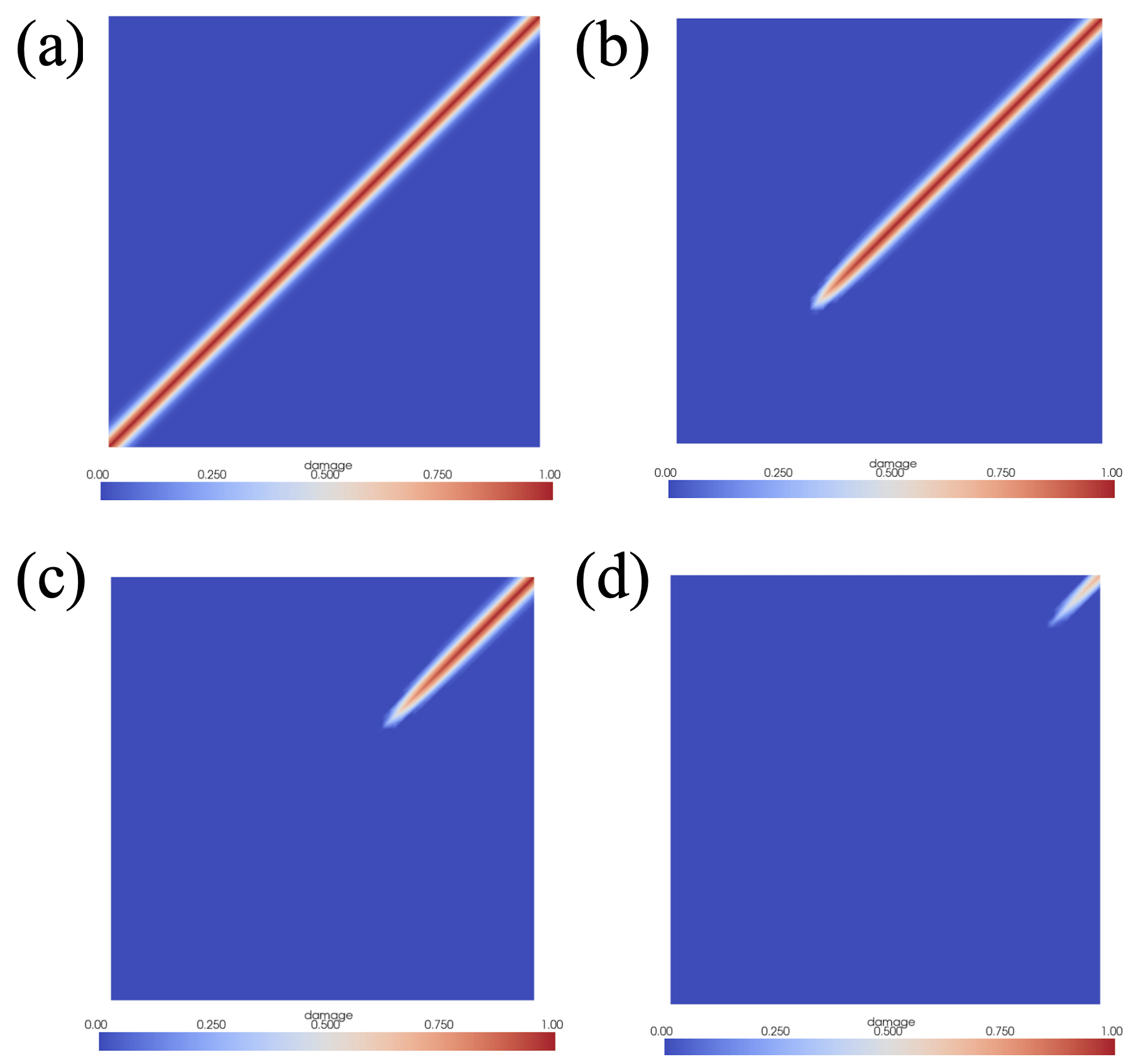}
    \caption{Healing Progress for $\beta = 45^\circ$ at four different representative times. (a) Beginning of the simulation, (b) $\sim$1/3 of the simulation time, (c) $\sim$2/3 of the simulation time, (d) Final time.}
    \label{fig:healing_stages_beta_pi_4}
\end{figure}

\begin{figure}[htp]
    \centering
    \includegraphics[width=0.8\linewidth]{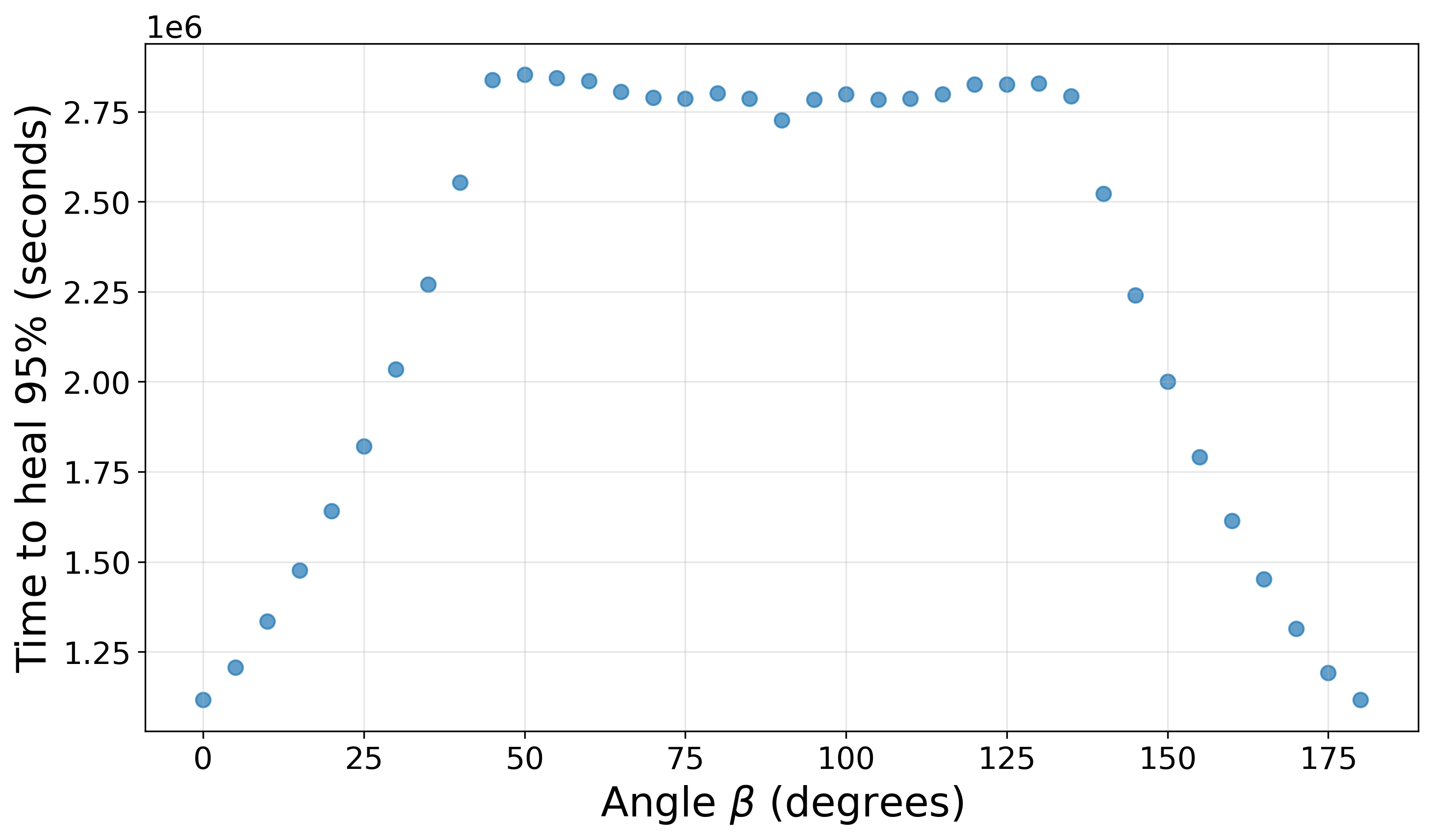}
    \caption{Time to Heal 95\% (seconds) vs. Crack Angle $\beta$ (degrees)}
    \label{fig:healing_time_vs_crack_angle}
\end{figure}

The impact of the crack angle on the healing time was investigated using the CDM. As an example, Figure~\ref{fig:healing_stages_beta_pi_4} shows the healing progress for $\beta = 45^\circ$ at four different representative moments in time. The healing time is defined as the time to heal 95\% of the volume determined by the damage surface integral. Within the range of $\beta \in [0^\circ, 180^\circ]$, 37 values were chosen and simulated. Other parameters were kept constant. Specifically, their values were $\alpha$ = 0.01, $\sigma$ = 0.0224, and $\gamma$ = 0.0316. The healing time versus crack angle plot is shown in Figure~\ref{fig:healing_time_vs_crack_angle}. 

The model simulation provided important physical insights. Between $\beta = 0^\circ$ and $45^\circ$, the scatterplot exhibited a convex function with positive slope. As $\beta$ increased within this range, water required greater diffusion distance to reach the rightmost crack endpoint, which approached the right boundary of the unit square. Due to symmetry about $\beta = 90^\circ$ in the range $\beta = [0^\circ, 180^\circ]$, this behavior was mirrored in the region from $\beta = 135^\circ$ to $180^\circ$. The middle region displayed different healing behavior, where healing times remained relatively constant because the rightmost point maintained the same distance from the initial moisture source (1 cm), requiring water to travel identical distances for $\beta = 45^\circ$ to $135^\circ$. As $\beta$ approached $90^\circ$ (horizontal), healing time decreased slightly. Since the diffusion coefficient of cracked regions $D_{\text{cracked}}$ exceeded that of intact concrete $D_{\text{intact}}$, water transport through cracks was faster than water transport through concrete. Crack length varied with orientation angle $\beta$, resulting in minimum healing time at $\beta = 90^\circ$ (shortest horizontal crack) and maximum healing times at $\beta = 45^\circ$ and $135^\circ$ (longest diagonal cracks). Ultimately, diffusion distance was the primary factor determining the total healing time.

\begin{figure}[htp]
    \centering
    \includegraphics[width=0.8\linewidth]{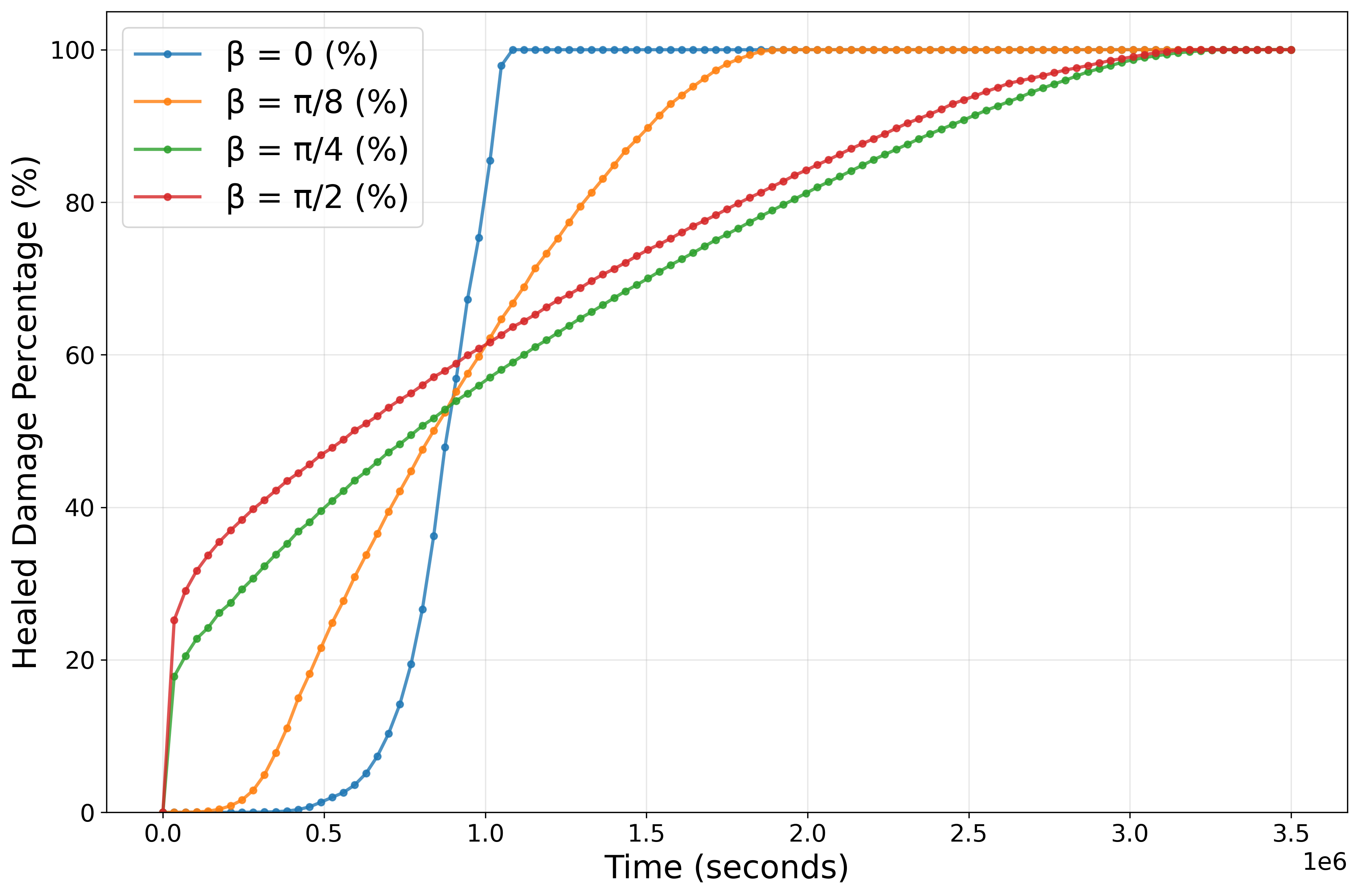}
    \caption{Comparison of Healing Progress for Selected Crack Angles}
    \label{fig:combined_healing_progress}
\end{figure}

To examine the complexity of healing behavior across different crack angles $\beta$, healing progression plots were generated for $\beta = 0$, $\frac{\pi}{8}$, $\frac{\pi}{4}$, and $\frac{\pi}{2}$, as shown in Figure~\ref{fig:combined_healing_progress}. Comparison between $\beta = 0$ and $\beta = \frac{\pi}{8}$ reveals that the crack at angle $\beta = \frac{\pi}{8}$ initiated healing earlier than the vertical crack ($\beta = 0$), as water reached the leftmost crack region sooner for $\beta = \frac{\pi}{8}$. However, the healing rate remained relatively slow because water had to continue saturating the domain until reaching the rightmost crack region of $\beta = \frac{\pi}{8}$. In contrast, once the vertical crack ($\beta = 0$) began healing, it healed at a significantly faster rate because water could immediately begin traversing the entire crack width. Consequently, the vertical crack achieved complete healing in a shorter time due to reduced diffusion distance. For $\beta = 0$ and $\frac{\pi}{8}$, the healing exhibited a symmetrical sigmoid shape. Comparison between the $\beta = \frac{\pi}{4}$ and $\beta = \frac{\pi}{2}$ cracks indicates that the horizontal crack ($\beta = \frac{\pi}{2}$) healed slightly faster than the $\beta = \frac{\pi}{4}$ crack. This observation is consistent with the Healing Time vs. Crack Angle plot in Figure~\ref{fig:healing_time_vs_crack_angle}, which demonstrates the influence of crack length on healing time when the crack mouth is in direct contact with the water source. Examination of the $\beta = \frac{\pi}{4}$ and $\frac{\pi}{2}$ graphs reveals rapid damage recovery at the beginning, then steadily increasing healing progress in the middle, and finally slowed healing progression as total healing approached 100\%. Therefore, the 95\% healed damage threshold was necessary; otherwise, healing times for full damage recovery would have been substantially longer, distorting the results. All graphs showed a 50\% healing threshold at approximately $8 \times 10^5$ seconds, representing the time required for water to reach the crack center. This observation further supports the conclusion that healing time is primarily determined by diffusion processes. 

Crack orientation has not been analyzed as a factor in diffusion-based autogenous healing models. Previous studies typically assume cracks are perpendicular to the diffusion direction or that the crack mouth is in direct contact with the water source. This work examines healing behavior across a range of crack orientations. 

\subsection{Healing Time vs. Crack Width}
\begin{figure}[htp]
    \centering    
    \includegraphics[width=1\textwidth]{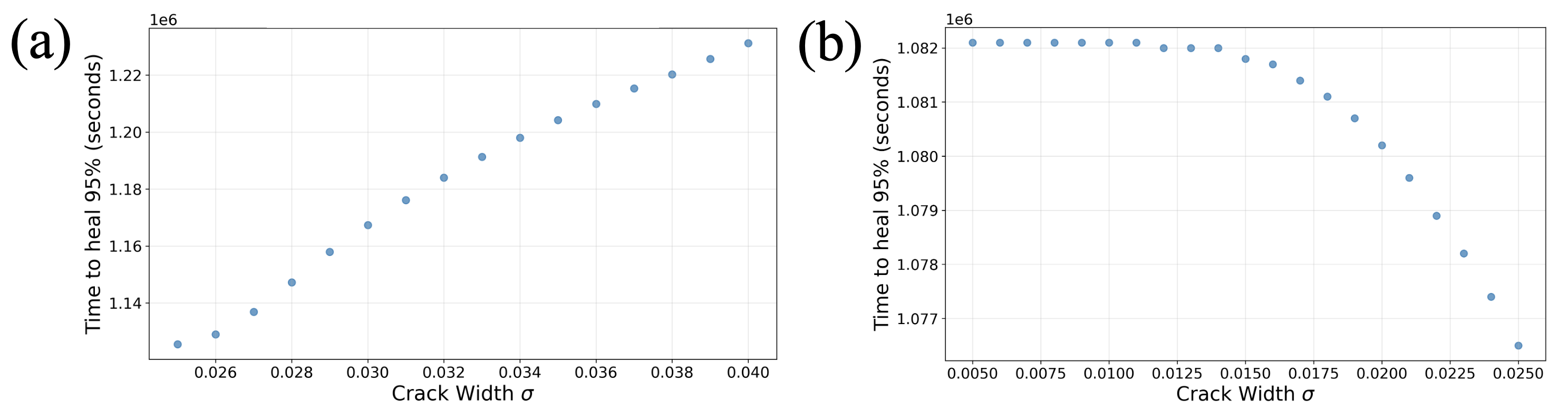}
    \caption{Time to Heal 95\% vs. Crack Width Parameter $\sigma$ (cm): (a) Simulation 1, (b) Simulation 2}
    \label{fig:healing_width_comparison}
\end{figure}

The impact of the crack width on the healing time was investigated using the CDM. Two simulations were conducted with different parameter values. Within the range of $\sigma \in [0.005, 0.040]$ cm, 37 values were chosen and simulated. For Simulation 1, its parameter values were $D_{\text{intact}}=10^{-8}$ cm$^2$/s, $D_{\text{cracked}}=10^{-9}$ cm$^2$/s, $\alpha$ = 0.01, $\beta$ = 0, and $\gamma$ = 0.0316. For Simulation 2, its parameter values were $D_{\text{intact}}=10^{-8}$ cm$^2$/s, $D_{\text{cracked}}=10^{-7}$ cm$^2$/s, $\alpha$ = 0.01, $\beta$ = 0, and $\gamma$ = 0.0316. The healing time versus crack width plots are shown in Figure~\ref{fig:healing_width_comparison}. 

Simulation 1 exhibited a steadily increasing healing time as crack width increased from $\sigma=0.025$ cm to $\sigma=0.040$ cm, with slight concavity. By contrast, Simulation 2 revealed a decreasing concave relationship between healing time and crack width on the interval $\sigma=0.005$ cm to $\sigma=0.025$ cm. In Simulation 1, the diffusion coefficient in cracked regions was less than that in intact regions, resulting in increased healing time as crack width increased. In Simulation 2, the diffusion coefficient in cracked regions exceeded that in intact regions, enabling rapid diffusion and healing processes across the crack interface. Since the healing process is diffusion-limited, larger crack widths reduce the effective diffusion distance, thereby decreasing the time required for water transport and allowing accelerated healing. The healing time increases with crack width when crack diffusivity is less than concrete diffusivity, whereas healing time decreases with increasing crack width when crack diffusivity exceeds concrete diffusivity.

\begin{figure}[htp]
    \centering
    \includegraphics[width=\linewidth]{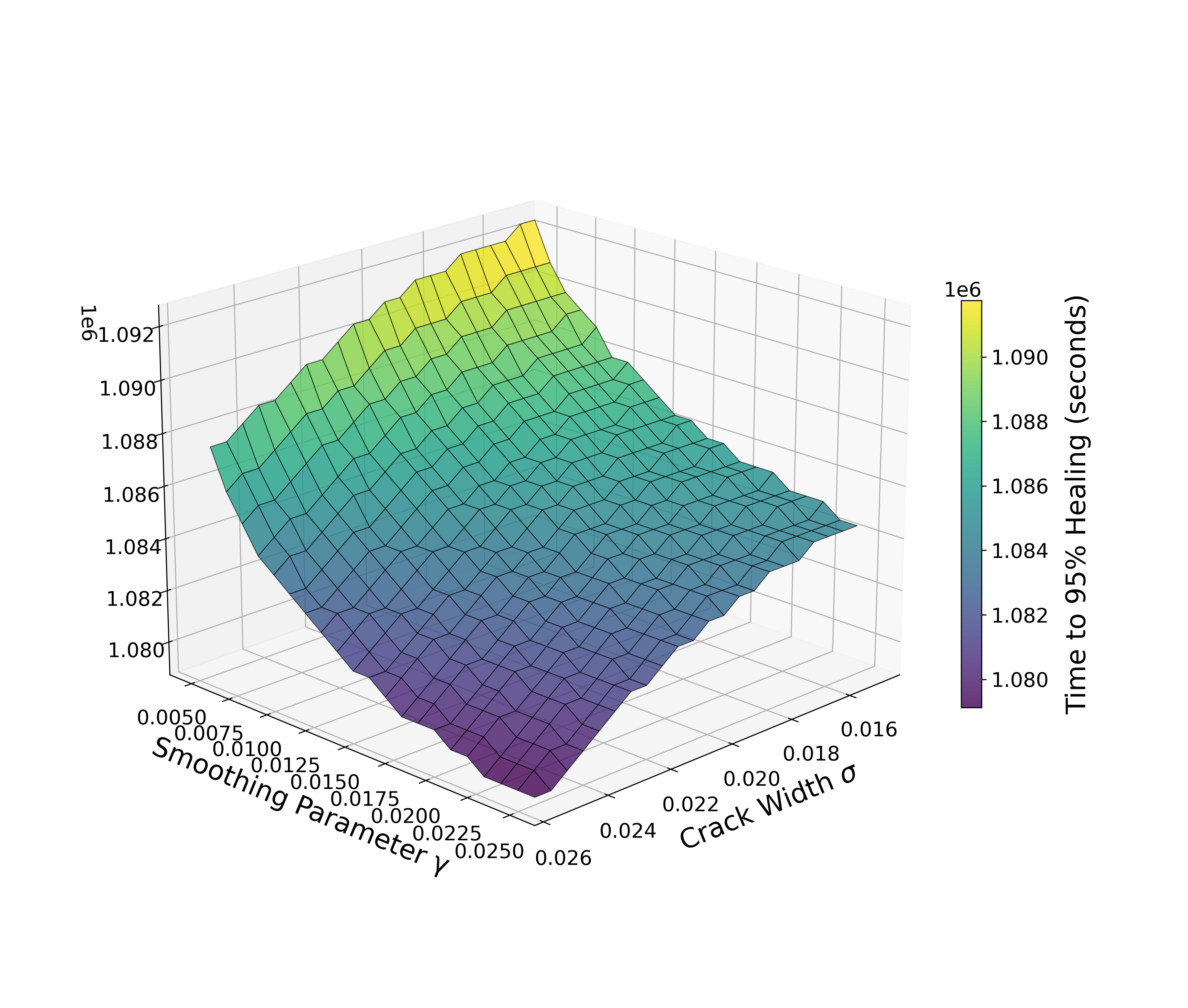}
    \caption{3D Plot of Healing Time $t$, Crack Width $\sigma$, and Smoothing Parameter $\gamma$}
    \label{fig:3d_surface_plot}
\end{figure}

The smoothing parameter $\gamma$ is closely related to the crack width indicator $\sigma$ and governs the spatial extent over which cement clinker can be transported and solidified. The healing time exhibits a functional dependence on both $\sigma$ and $\gamma$ when other parameters remain constant. Figure~\ref{fig:3d_surface_plot} presents this functional relationship as a three-dimensional surface plot. The analysis reveals that healing time increases as the smoothing parameter $\gamma$ decreases. This relationship occurs because decreasing $\gamma$ reduces the effective distance over which cement hydration can occur from its source. The healing time demonstrates a decreasing convex relationship with the smoothing parameter, indicating diminishing effects from increasing $\gamma$. Conversely, the healing time exhibits a decreasing concave relationship with crack width, displaying accelerated time effects as crack width increases. The functional shapes of $\sigma$ and $\gamma$ remain relatively stable when one parameter is varied while the other is held constant. 

Previous studies have shown that the healing process is highly dependent on crack width \cite{ruan2020influence}. For wider cracks, healing has been shown to be diffusion-limited \cite{vedrtnam2025bacterial}, consistent with our finding that increasing crack width decreases the diffusion distance and time to the crack boundary. The cement smoothing parameter introduced here has not been implemented in previous models. 

\subsection{Healing Percentage vs. Time}
\begin{figure}[htp]
    \centering
    \includegraphics[width=0.8\linewidth]{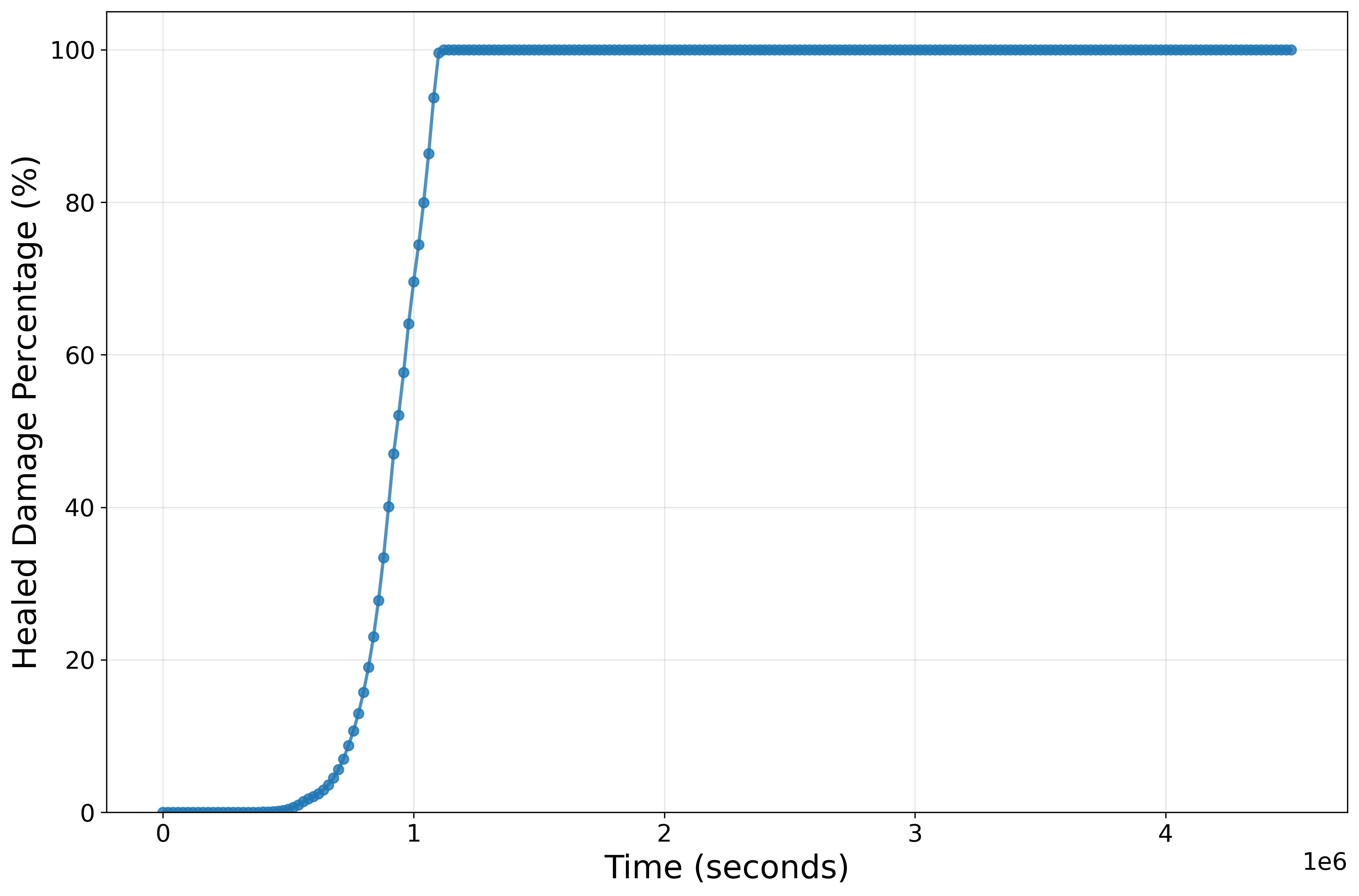}
    \caption{Healing Percentage (\%) as a Function of Time (seconds)}
    \label{fig:CDM_healing_percentage_over_time}
\end{figure}

The healing process is also of interest for practical use. Figure~\ref{fig:CDM_healing_percentage_over_time} illustrates the typical healing progression over time simulated using the CDM with parameter values $\alpha$ = 0.01, $\beta$ = 0, $\sigma$ = 0.0224, and $\gamma$ = 0.0316. The healing process demonstrated an initial slow phase followed by acceleration as water saturated the entire crack. The healing rate approached an asymptotic maximum until complete crack healing was achieved. This behavior reflects the rate-limiting nature of water diffusion in the healing process. As damage decreased, water diffusion efficiency increased, resulting in higher healing rates. Similar healing patterns were observed across all parameter configurations with the CDM. Previous studies have demonstrated that crack closure does not occur under low-humidity conditions. In contrast, under wet-dry cycling conditions and sustained immersion, healing efficiency exhibits a progressive increase over time until reaching a maximum value \cite{roigflores2020concrete}. These observations are consistent with our model predictions, which indicate that elevated water saturation levels correspond to enhanced healing rates. 

\subsection{Comparison between CDM and CMM}
\begin{figure}[htp]
    \centering
    \includegraphics[width=0.8\linewidth]{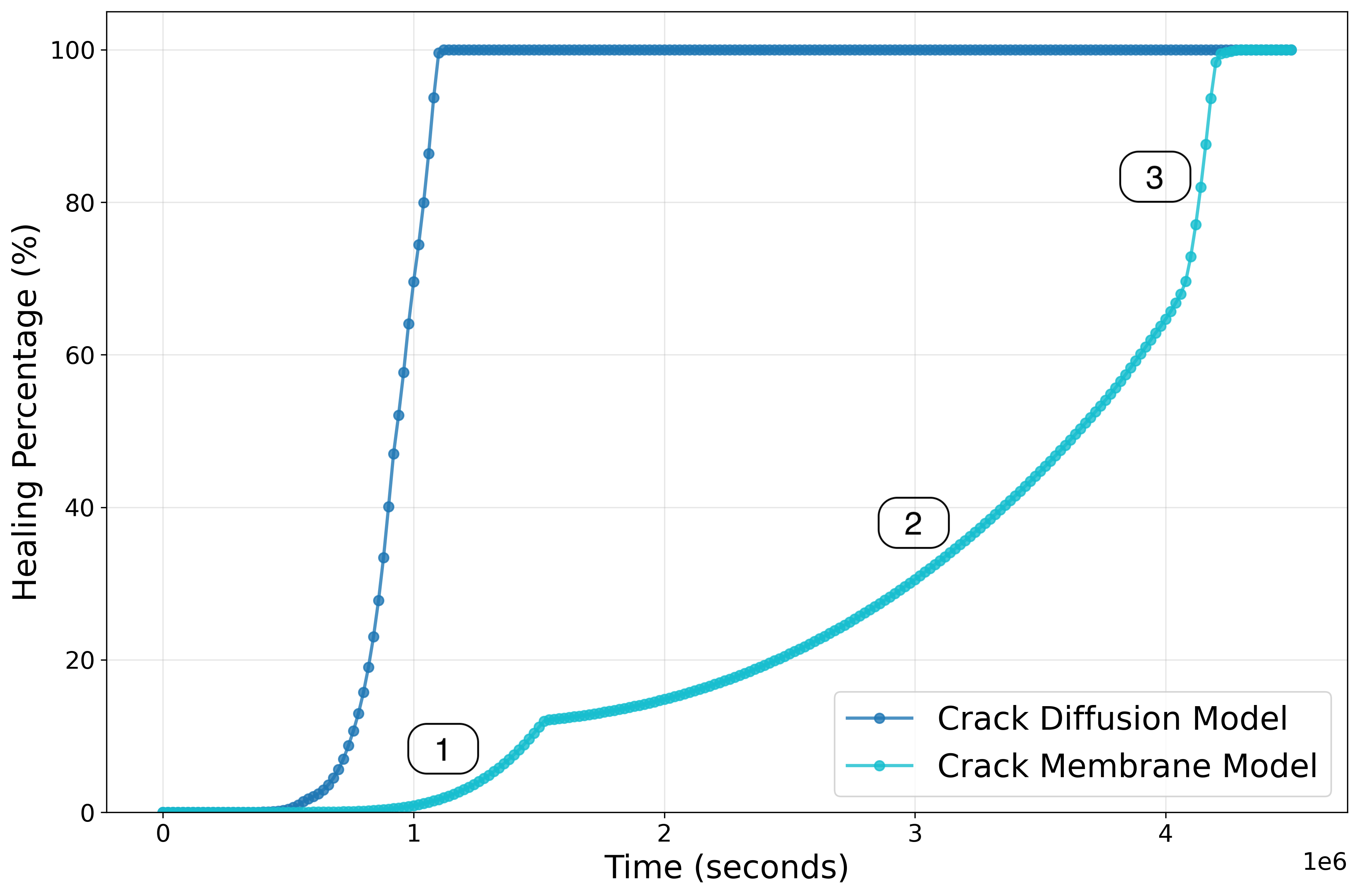}
    \caption{Crack Diffusion vs. Crack Membrane Model Healing Percentage (\%) as a Function of Time (seconds)}
    \label{fig:combined_model_plotting}
\end{figure}

The healing progressions of the two models, CDM and CMM, were contrasted, with parameter values set to $\alpha$ = 0.01, $\beta$ = 0, $\sigma$ = 0.0224, and $\gamma$ = 0.0316. The results are presented in Figure~\ref{fig:combined_model_plotting}.

\begin{figure}[htp]
    \centering
    \includegraphics[width=0.8\linewidth]{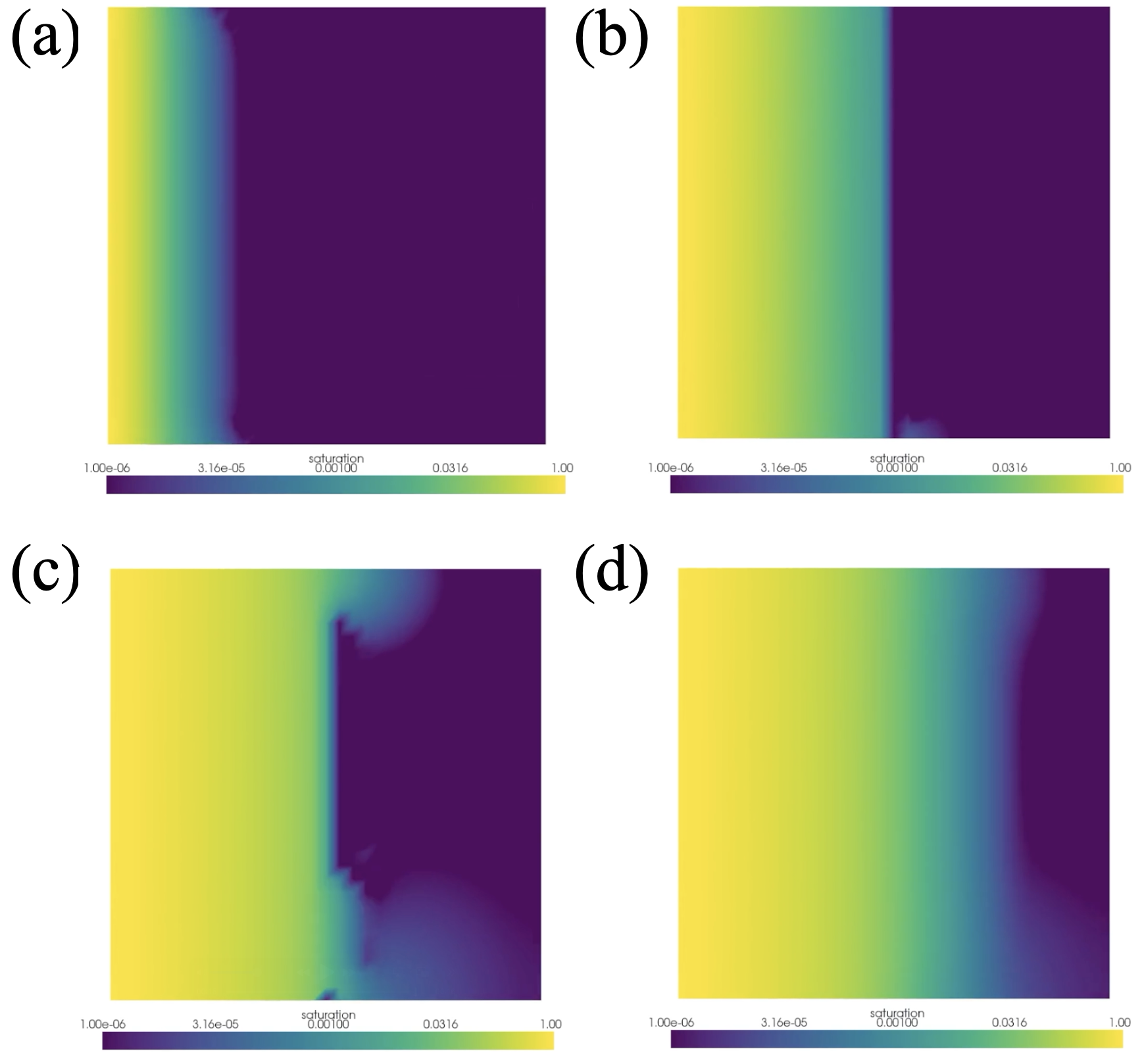}
    \caption{Saturation Field of the Crack Membrane Model at Four Different Representative Times. (a) Diffusion Toward the Crack, (b) Water Flow Blocked at the Crack Membrane, (c) Partial Water Release, (d) Full Water Release.}
    \label{fig:CMM_saturation_stages}
\end{figure}

The analysis revealed a unique healing process with the CMM. There are three distinctive stages. Stage 1 represents the preliminary healing of partially damaged regions as water diffuses into these areas. Once water reaches the crack, it is stopped at the interface to prevent further movement. This is represented as the end of Stage 1 and a sharp corner in the graph. When further rightward water movement is suspended, Stage 2 of the healing process returns to the rate characteristic of healing within the already saturated regions. The end of Stage 2 occurs when the moisture threshold is reached, and the membrane begins releasing the water held at the left side of the crack boundary. Then, it enters the final phase, Stage 3, when water rapidly saturates the rest of the crack and heals the remainder. In Stage 3, the healing rate reaches a maximum. While the corners of this graph are quite dramatic, they have been emphasized through the crack gate sharpness parameter for the purposes of this illustration. To demonstrate these mechanisms, Figure~\ref{fig:CMM_saturation_stages} shows the saturation field progression of the Crack Membrane Model.

Comparing the two models in Figure~\ref{fig:combined_model_plotting}, the added functionality of halting water movement until sufficient moisture is met in the Crack Membrane Model significantly lengthens the healing time. In the beginning (during Stage 1), the Crack Membrane Model exhibits a substantially slower initial healing rate compared to the Crack Diffusion Model. During Stage 3, the Crack Membrane Model achieves healing rates comparable to the final healing rates of the Crack Diffusion Model. The Crack Membrane Model provides a more accurate representation of the underlying physical and chemical processes. Experimental validation should demonstrate these characteristic transitions and distinct healing stages. 

Previous studies have shown that standard linear diffusion approaches are inadequate for cracks under viscous or inertial pressure gradients. For these cases, crack width and roughness determine the nonlinear flow of water \cite{hou2024exploring}. While past models have documented pressure-based water flow through cracks, the present work applies the mechanism to self-healing concrete.

The Crack Membrane Model's additional calibration requirements increase analytical complexity. Thus, the Crack Diffusion Model was chosen as the primary model for the other analyses. 

\subsection{Performance Comparison of Machine Learning Models}

\begin{figure}[htp]
    \centering
    \includegraphics[width=1\linewidth]{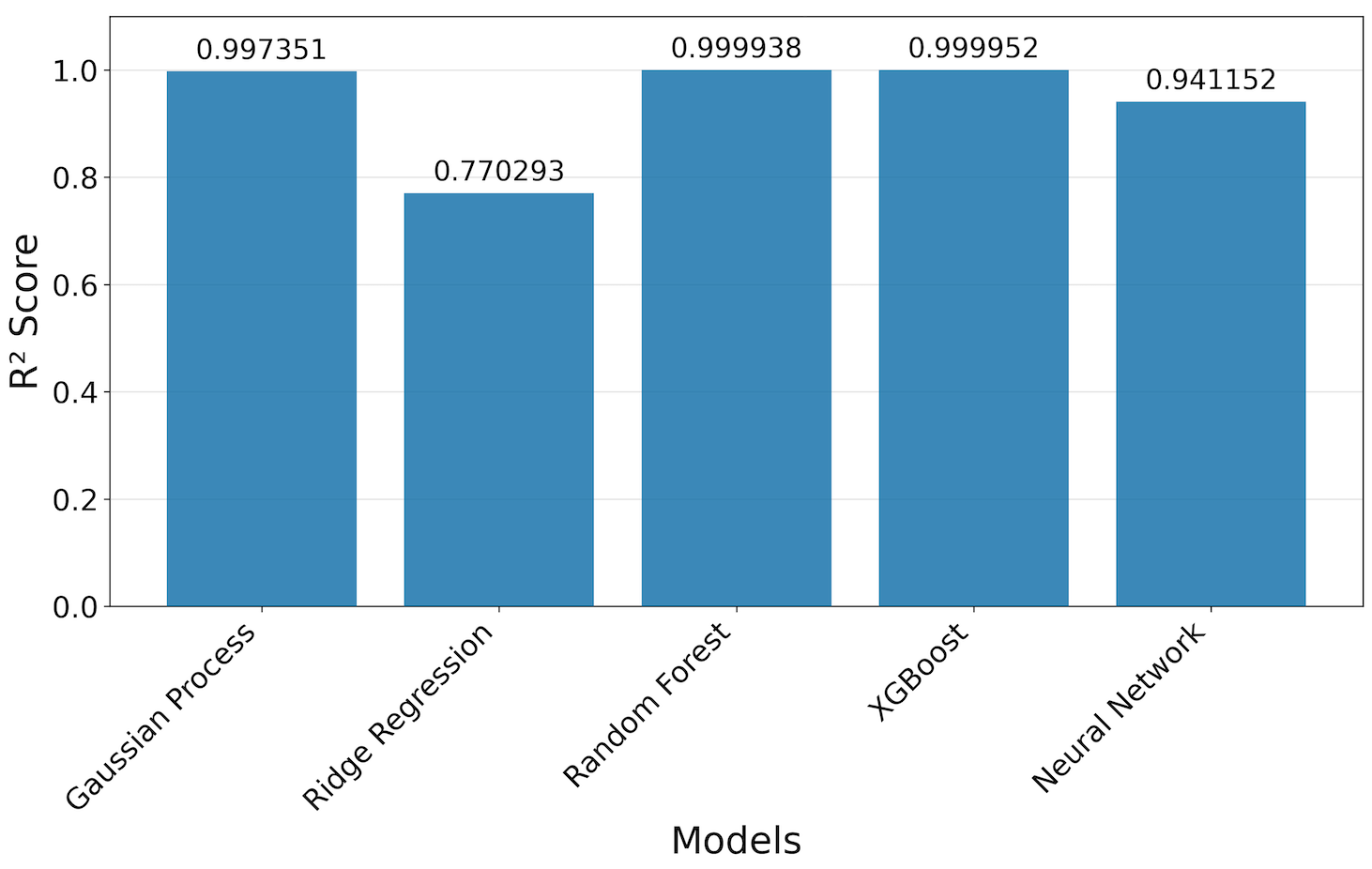}
    \caption{Machine Learning Model Performance Comparison Bar Chart ($R^2$)}
    \label{fig:ml_model_performance_comparison}
\end{figure}
The $R^2$ score performance comparison for the five machine learning models is shown in Figure~\ref{fig:ml_model_performance_comparison}. The Gradient Boosting Model (XGBoost) achieved the highest performance with an $R^2$ score of 0.999952, slightly higher than the Random Forest Model with an $R^2$ score of 0.999938. The performances of the Gaussian Process, Neural Network, and Ridge Regression Models followed, in that order. Previous studies have demonstrated that regression-based machine learning models can predict self-healing behavior in concrete with $R^2$ values typically exceeding $0.9$ \cite{chiadighikaobi2024predicting}. The present model achieves an $R^2$ value of $0.999952$, which surpasses these prior benchmarks and demonstrates sufficient accuracy to serve as an effective surrogate for finite element simulations. 

\begin{figure}[htp]
    \centering
    \includegraphics[width=1\linewidth]{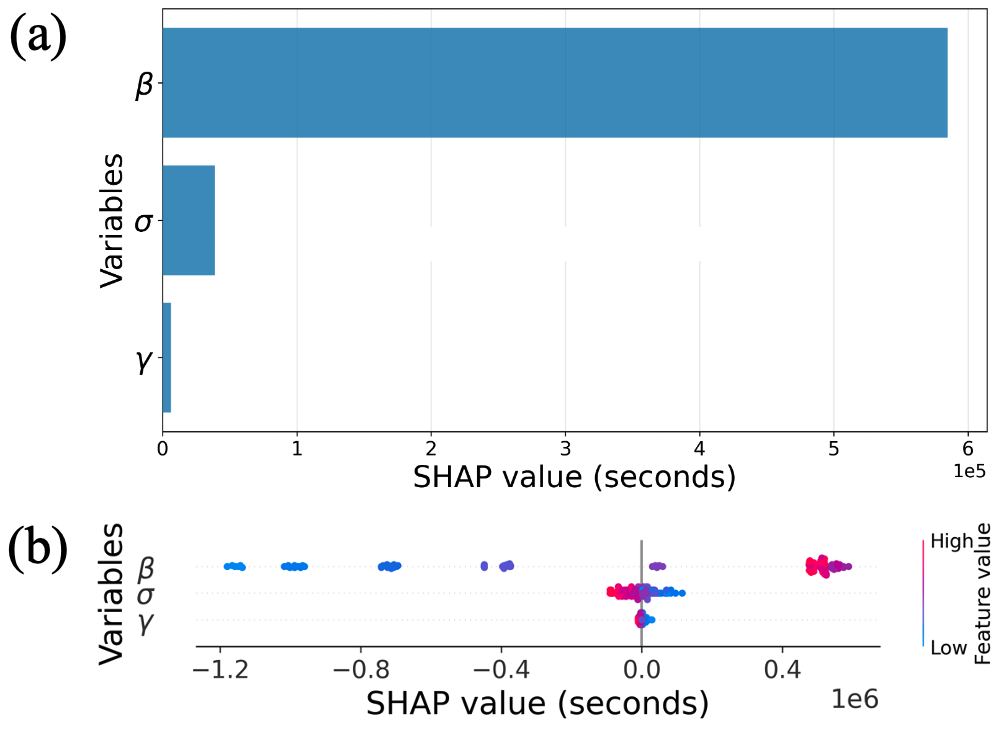}
    \caption{SHAP Analysis of XGBoost Machine Learning Model (a) Mean Absolute SHAP Value Comparison Bar Chart, (b) SHAP Feature Impact Distribution Summary Plot}
    \label{fig:shap_analysis}
\end{figure}

SHAP was used to analyze the XGBoost model. The analysis results are shown in Figure~\ref{fig:shap_analysis}. The mean absolute SHAP value comparison bar chart (Figure~\ref{fig:shap_analysis}(a)) quantifies the average magnitude of each feature's contribution, revealing that crack orientation $\beta$ exhibits the highest mean absolute impact on healing time predictions at $5.85 \times 10^5$ seconds, followed by crack width indicator $\sigma$ at $3.88 \times 10^4$ seconds and cement availability smoothing parameter $\gamma$ at $6.21 \times 10^3$ seconds. This observation is consistent with experimental findings reporting that healing efficiency and rate are strongly governed by the initial damage distribution \cite{pan2018damage}. Because crack orientation $\beta$ and crack width indicator $\sigma$ exert the largest and second-largest influence on the damage distribution, respectively, it is reasonable that they also exert the largest and second-largest influence on the healing time.

The SHAP feature impact distribution summary plot (Figure~\ref{fig:shap_analysis}(b)) provides detailed insights into the directional and value-dependent influence of each feature. Higher values of $\beta$ (in red) tend to increase the predicted healing time (positive SHAP values), while lower values (in blue) tend to decrease the predicted healing time (negative SHAP values). The opposite is observed for $\sigma$ and $\gamma$. 

Crack orientation $\beta$ demonstrates a predominantly left-skewed SHAP value distribution with lower values sparsely distributed on the negative side, while higher values are densely concentrated on the positive side. This reinforces the notion that the healing time is diffusion-time limited, since the diffusion distance to the crack endpoint for $\beta = 45^{\circ}$ to $\beta = 90^{\circ}$ is approximately constant at $1$ cm. 

Crack width indicator $\sigma$ exhibits an approximately symmetric range centered near zero, but shows a large cluster of negative SHAP values for higher values of $\sigma$. This relationship aligns with Figure~\ref{fig:healing_width_comparison} and Figure~\ref{fig:3d_surface_plot}, where higher $\sigma$ values result in a decrease in the healing time. As $\sigma$ continues to increase, the decrease in healing time accelerates, causing a slight right-skew in the SHAP value distribution.

Cement availability smoothing parameter $\gamma$ displays a strong right-skewed SHAP value distribution. As $\gamma$ increases, the healing time decreases. This relationship is consistent with Figure~\ref{fig:3d_surface_plot}, which shows that healing time exhibits a convex, decreasing dependence on $\gamma$.

\subsection{Limitations and Future Improvements}\label{sec:limitations_and_future_extensions}

Several limitations constrain the current model's applicability. First, model parameters were estimated rather than experimentally calibrated, potentially affecting quantitative predictions while preserving qualitative trends. Second, crack geometries were simplified to Gaussian distributions, whereas real cracks exhibit complex propagation patterns that could be better captured through techniques such as X-ray microtomography. Third, the model assumes constant cement availability in newly healed regions, which may overestimate long-term healing capacity. Fourth, water transport is modeled through diffusion rather than pressure-driven flow governed by porous media laws such as Darcy's law.

Future work should address these limitations through experimental validation, realistic crack geometry implementation, and incorporation of porous media flow mechanics. Validation against experimental data is essential to verify simulation reliability. This validation should include direct comparison of predicted healing times and damage evolution patterns with laboratory measurements. Mesh convergence studies and comparison with analytical solutions will further establish the numerical accuracy of the finite element analysis \cite{asme2021vv20}. The models could be extended to three-dimensional domains to capture volumetric healing effects, cyclic loading scenarios to assess long-term durability, and strength recovery calculations through post-processing of damage fields. Incorporation of additional healing factors, including concrete hydration age, environmental curing conditions, curing history, and evolving microstructure, would further enhance the model's predictive capability for real-world applications. The machine learning models can be further optimized using physics informed neural networks (PINNs), incorporating a loss function that penalizes the model's predictions if it violates the diffusion and healing processes.

\section{Conclusions}\label{sec:conclusions}

This study developed two coupled finite element models for simulating autogenous self-healing concrete behavior, addressing the critical need for computational tools to evaluate this promising technology for infrastructure reenforcement and sustainability. The Crack Diffusion Model (CDM) and Crack Membrane Model (CMM) successfully integrated time-dependent feedback mechanisms, logarithmic interpolation of diffusion coefficients, and damage evolution based on cement and water availability. The models implement a Helmholtz filtering approach to model the distance over which cement clinker can travel to form crystals during the healing process. CMM adds a threshold-based binary gate mechanism to the CDM to simulate the moisture threshold buildup and rapid healing after gate activation. This study developed a machine learning framework that serves as a more computationally efficient surrogate for the finite element models. 

The models revealed several key insights into healing behavior:
\begin{itemize}
\item Crack orientation significantly influences healing time, with diffusion distance to crack endpoints being the dominant factor rather than crack length. Healing time increases from $0^\circ$ to $45^\circ$ as water travels farther to reach crack endpoints, then remains relatively constant from $45^\circ$ to $135^\circ$ due to similar diffusion distances, with a slight decrease at $90^\circ$ due to shorter crack length.
\item The relationship between crack width and healing time depends on the relative diffusivities of cracked versus intact regions. If the diffusion coefficient in cracked regions is less than that in intact regions, the healing time increases with crack width. If the diffusion coefficient in cracked regions is greater than that in intact regions, the healing time decreases with crack width. 
\item A 3D surface plot revealing the intertwined influence of crack width and cement availability smoothing on healing time showed that healing time decreases as the smoothing parameter increases. Changing the smoothing parameter preserved the shape of the healing time versus crack width function, but shifted it along a convex, decreasing curve. 
\item The CMM provides a more realistic representation of healing physics through its binary gate mechanism, showing three distinct healing phases: initial diffusion, moisture buildup, and rapid healing after gate activation. However, the CDM offers computational efficiency advantages for parametric studies.
\item Machine learning models trained on over one thousand simulation data points achieved high accuracy (up to $R^2=0.999952$ for the Gradient Boosting Model) in predicting healing time, demonstrating the potential for rapid screening of design parameters. 
\item SHAP analysis demonstrated that crack orientation influenced the healing time the most, followed by crack width and cement availability smoothing. This demonstrates that the healing time is diffusion-time limited, since angle and width of the crack are the most important factors in determining the diffusion distance. The feature impact distribution summary plot aligned with results from the finite element analysis. 
\end{itemize}

These findings contribute to the understanding of autogenous healing mechanisms and provide a foundation for optimizing self-healing concrete formulations. The models address limitations in existing literature by incorporating time-dependent feedback, realistic cement availability smoothing, and machine learning integration for rapid prediction. While experimental validation remains necessary for quantitative accuracy, the models offer valuable insights for guiding future laboratory studies and field applications.

Additional factors that can be modeled include concrete hydration age, environmental curing conditions, curing history, and evolving microstructure. Future work should focus on experimental calibration, three-dimensional extensions, and cyclic loading scenarios to further enhance predictive capability for real-world applications. 
 
The computational framework developed in this study could accelerate the implementation of self-healing concrete technologies by enabling rapid parameter screening and optimization before expensive experimental trials, providing guidance for material design, and reducing the time and cost required to evaluate healing performance across diverse conditions. The machine learning surrogate models further enhance this capability by delivering near-instantaneous predictions. This combined finite element and machine learning approach reduces the risk and uncertainty associated with field implementation, thereby accelerating the adoption of self-healing concrete, which has the potential to reduce the environmental and economic costs associated with infrastructure maintenance and repair. 

\section*{CRediT Author Contribution Statement}
\textbf{William Liu}: Conceptualization, Data Curation, Formal Analysis, Investigation, Methodology, Project Administration, Resources, Software, Validation, Visualization, Writing - Original Draft, Writing - Review and Editing.

\section*{Declaration of Competing Interest}
The author declares that they have no known competing financial interests or personal relationships that could have appeared to influence the work reported in this paper.

\section*{Acknowledgments}
The author is grateful for the guidance and insights provided by Natalie Ronco from Cambridge University. 

\section*{Data Availability}
A permanent copy of code, data, and outputs of this work are publicly available at \url{https://github.com/william-liu0/concrete-healing-model}.


\begin{thebibliography}{99}
\expandafter\ifx\csname url\endcsname\relax
  \def\url#1{\texttt{#1}}\fi
\expandafter\ifx\csname urlprefix\endcsname\relax\def\urlprefix{URL }\fi
\expandafter\ifx\csname href\endcsname\relax
  \def\href#1#2{#2} \def\path#1{#1}\fi

\bibitem{nguyen2023review}
M.-T. Nguyen, C.~A. Fernandez, M.~M. Haider, K.-H. Chu, G.~Jian, S.~Nassiri,
  D.~Zhang, R.~Rousseau, V.-A. Glezakou, Toward {{Self-Healing Concrete
  Infrastructure}}: {{Review}} of {{Experiments}} and {{Simulations}} across
  {{Scales}}, Chemical Reviews 123~(18) (2023) 10838--10876.
\newblock \href {https://doi.org/10.1021/acs.chemrev.2c00709}
  {\path{doi:10.1021/acs.chemrev.2c00709}}.

\bibitem{nachi2025history}
N.~G.~K. Shepard, The history of concrete,
  https://www.nachi.org/history-of-concrete.htm, accessed 30 July 2025 (2025).

\bibitem{gitnux2025concrete}
J.~Linder, Concrete industry statistics: Market data report 2025,
  https://gitnux.org/concrete-industry-statistics/, accessed 30 July 2025
  (2025).

\bibitem{andrewGlobalCO2Emissions2024}
R.~M. Andrew, Global {{CO2}} emissions from cement production (May 2024).
\newblock \href {https://doi.org/10.5281/zenodo.11207133}
  {\path{doi:10.5281/zenodo.11207133}}.

\bibitem{weforum2024cement}
M.~Purton, Cement production sustainable concrete co2 emissions,
  https://www.weforum.org/stories/2024/09/cement-production-sustainable-concrete-co2-emissions/,
  accessed 30 July 2025 (2024).

\bibitem{rmi2023concrete}
B.~Skinner, R.~Lalit, With concrete, less is more,
  https://rmi.org/with-concrete-less-is-more/, accessed 30 July 2025 (January
  2023).

\bibitem{structuremag2016durability}
P.~Noyce, G.~Crevello, Durability of reinforced concrete,
  https://www.structuremag.org/article/durability-of-reinforced-concrete/,
  accessed 30 July 2025 (January 2016).

\bibitem{asce2025infrastructure}
{American Society of Civil Engineers}, Asce's 2025 infrastructure report card,
  https://infrastructurereportcard.org/, accessed 30 July 2025 (2025).

\bibitem{han2017smart}
B.~Han, L.~Zhang, J.~Ou, Smart and Multifunctional Concrete Toward Sustainable
  Infrastructures, Springer, Singapore, 2017.

\bibitem{zhang2020selfhealing}
W.~Zhang, Q.~Zheng, A.~Ashour, B.~Han, Self-healing cement concrete composites
  for resilient infrastructures: A review, Composites Part B: Engineering 189
  (2020) 107892.
\newblock \href {https://doi.org/10.1016/j.compositesb.2020.107892}
  {\path{doi:10.1016/j.compositesb.2020.107892}}.

\bibitem{nace2025corrosion}
N.~International, Global cost of corrosion,
  http://impact.nace.org/economic-impact.aspx, accessed 30 July 2025 (2025).

\bibitem{hearnSelfsealingAutogenousHealing1998}
N.~Hearn, Self-sealing, autogenous healing and continued hydration: {{What}} is
  the difference?, Materials and Structures 31~(8) (1998) 563--567.
\newblock \href {https://doi.org/10.1007/BF02481539}
  {\path{doi:10.1007/BF02481539}}.

\bibitem{edvardsenWaterPermeabilityAutogenous1999a}
C.~Edvardsen, Water {{Permeability}} and {{Autogenous Healing}} of {{Cracks}}
  in {{Concrete}}, Materials Journal 96~(4) (1999) 448--454.
\newblock \href {https://doi.org/10.14359/645} {\path{doi:10.14359/645}}.

\bibitem{amran2022selfhealing}
M.~Amran, A.~M. Onaizi, R.~Fediuk, N.~I. Vatin, R.~S.~M. Rashid, H.~Abdelgader,
  T.~Ozbakkaloglu, Self-healing concrete as a prospective construction
  material: A review, Materials 15~(9) (2022) 3214.
\newblock \href {https://doi.org/10.3390/ma15093214}
  {\path{doi:10.3390/ma15093214}}.

\bibitem{usc2022bacteria}
{USC Viterbi School of Engineering}, Can concrete heal its own cracks without
  losing its strength?,
  https://viterbischool.usc.edu/news/2022/03/can-concrete-heal-its-own-cracks-without-losing-its-strength/,
  accessed 30 July 2025 (2022).

\bibitem{lahmann2022autogenous}
D.~Lahmann, C.~Edvardsen, S.~Kessler, Autogenous self-healing of concrete:
  Experimental design and test methods---a review, Engineering Reports 5~(1)
  (2022) e12565.
\newblock \href {https://doi.org/10.1002/eng2.12565}
  {\path{doi:10.1002/eng2.12565}}.

\bibitem{bbc2015selfhealing}
BBC, 'self-healing' concrete trial launched in south wales,
  \url{https://www.bbc.com/news/uk-wales-south-east-wales-34658038}, accessed
  30 July 2025 (2015).

\bibitem{fenics}
{{FEniCS}}, https://fenicsproject.org/, accessed 30 July 2025 (2025).

\bibitem{hilloulin2014autogenous}
B.~Hilloulin, F.~Grondin, M.~Matallah, A.~Loukili, Modelling of autogenous
  healing in ultra high performance concrete, Cement and Concrete Research
  61--62 (2014) 64--70.
\newblock \href {https://doi.org/10.1016/j.cemconres.2014.04.003}
  {\path{doi:10.1016/j.cemconres.2014.04.003}}.

\bibitem{vedrtnam2025bacterial}
A.~Vedrtnam, K.~Kalauni, M.~T. Palou, Finite element simulation of bacterial
  self-healing in concrete using microstructural transport and precipitation
  modeling, Scientific Reports 15 (2025) 15809.
\newblock \href {https://doi.org/10.1038/s41598-025-99844-6}
  {\path{doi:10.1038/s41598-025-99844-6}}.

\bibitem{shahsavari2016damage}
H.~Shahsavari, M.~Baghani, S.~Sohrabpour, R.~Naghdabadi, Continuum
  damage-healing constitutive modeling for concrete materials through stress
  spectral decomposition, International Journal of Damage Mechanics 25~(6)
  (2016) 900--918.
\newblock \href {https://doi.org/10.1177/1056789515616447}
  {\path{doi:10.1177/1056789515616447}}.

\bibitem{chen2021hydration}
Q.~Chen, X.~Liu, H.~Zhu, J.~W. Ju, X.~Yongjian, Z.~Jiang, Z.~Yan, Continuum
  damage-healing framework for the hydration induced self-healing of the
  cementitious composite, International Journal of Damage Mechanics 30~(5)
  (2021) 681--699.
\newblock \href {https://doi.org/10.1177/1056789520968037}
  {\path{doi:10.1177/1056789520968037}}.

\bibitem{pan2018damage}
Y.~Pan, F.~Tian, Z.~Zhong, A continuum damage-healing model of healing agents
  based self-healing materials, International Journal of Damage Mechanics
  27~(5) (2017) 754--778.
\newblock \href {https://doi.org/10.1177/1056789517702211}
  {\path{doi:10.1177/1056789517702211}}.

\bibitem{alkhuzai2023numerical}
K.~Alkhuzai, L.~{Di Sarno}, A.~Haredy, R.~Alahmadi, D.~Albuhairi, Numerical
  simulation of the performance of self-healing concrete in beam elements,
  Buildings 13~(3) (2023) 809.
\newblock \href {https://doi.org/10.3390/buildings13030809}
  {\path{doi:10.3390/buildings13030809}}.

\bibitem{freeman2019simulation}
B.~L. Freeman, T.~Jefferson, The simulation of transport processes in
  cementitious materials with embedded healing systems, International Journal
  for Numerical and Analytical Methods in Geomechanics 44~(2) (2019) 293--326.
\newblock \href {https://doi.org/10.1002/nag.3017}
  {\path{doi:10.1002/nag.3017}}.

\bibitem{mauludin2021computational}
L.~M. Mauludin, T.~Rabczuk, Computational modeling of fracture in capsule-based
  self-healing concrete: {{A}} 3d study, Frontiers of Structural and Civil
  Engineering 15 (2021) 1337--1346.
\newblock \href {https://doi.org/10.1007/s11709-021-0781-1}
  {\path{doi:10.1007/s11709-021-0781-1}}.

\bibitem{kanellopoulos2022selfhealing}
A.~Kanellopoulos, J.~{Norambuena-Contreras} (Eds.), Self-Healing Construction
  Materials: Fundamentals, Monitoring and Large Scale Applications, Engineering
  Materials and Processes, Springer Nature Switzerland AG, Cham, Switzerland,
  2022.
\newblock \href {https://doi.org/10.1007/978-3-030-86880-2}
  {\path{doi:10.1007/978-3-030-86880-2}}.

\bibitem{badarinath2021machine}
P.~V. Badarinath, M.~Chierichetti, F.~{Davoudi Kakhki}, A machine learning
  approach as a surrogate for a finite element analysis: Status of research and
  application to one dimensional systems, Sensors 21~(5) (2021) 1654.
\newblock \href {https://doi.org/10.3390/s21051654}
  {\path{doi:10.3390/s21051654}}.

\bibitem{nath2020mlfea}
R.~Nath, S.~Patil, A.~Kumar, Machine learning accelerated finite element
  analysis, Computational Mechanics 65~(3) (2020) 567--589.
\newblock \href {https://doi.org/10.1007/s00466-019-01789-9}
  {\path{doi:10.1007/s00466-019-01789-9}}.

\bibitem{ruan2020influence}
S.~Ruan, J.~Qiu, E.-H. Yang, C.~Unluer, Influence of crack width on the
  stiffness recovery and self-healing of reactive magnesia-based binders under
  co$_2$--h$_2$o conditioning, Construction and Building Materials 269 (2021)
  121360.
\newblock \href {https://doi.org/10.1016/j.conbuildmat.2020.121360}
  {\path{doi:10.1016/j.conbuildmat.2020.121360}}.

\bibitem{roigflores2020concrete}
M.~{Roig-Flores}, P.~Serna, Concrete early-age crack closing by autogenous
  healing, Sustainability 12~(11) (2020) 4476.
\newblock \href {https://doi.org/10.3390/su12114476}
  {\path{doi:10.3390/su12114476}}.

\bibitem{hou2024exploring}
S.~Hou, K.~Li, X.~Hu, C.~Shi, Exploring the nonlinear behavior of flow through
  cracked concrete by water permeability test, Cement and Concrete Composites
  150 (2024) 105557.
\newblock \href {https://doi.org/10.1016/j.cemconcomp.2024.105557}
  {\path{doi:10.1016/j.cemconcomp.2024.105557}}.

\bibitem{chiadighikaobi2024predicting}
P.~C. Chiadighikaobi, M.~Hematibahar, M.~Kharun, N.~A. Stashevskaya, K.~Camara,
  Predicting mechanical properties of self-healing concrete with trichoderma
  reesei fungus using machine learning, Cogent Engineering 11~(1) (2024)
  2307193.
\newblock \href {https://doi.org/10.1080/23311916.2024.2307193}
  {\path{doi:10.1080/23311916.2024.2307193}}.

\bibitem{asme2021vv20}
{ASME}, V\&v 20-2009 (r2021): Standard for verification and validation in
  computational fluid dynamics and heat transfer,
  https://www.asme.org/codes-standards/find-codes-standards/standard-for-verification-and-validation-in-computational-fluid-dynamics-and-heat-transfer,
  originally published 2009, reaffirmed 2021 (2021).

\end{thebibliography}
\end{document}